\def\frac#1#2{{\textstyle{#1\over#2}}}
\def\darr#1{\raise1.5ex\hbox{$\leftrightarrow$}\mkern-16.5mu #1}
\def\){\right)}
\def\({\left( }
\def\]{\right] }
\def\[{\left[}
\def\si{{}^1\kern-.14em {\rm S}_0}
\def\siii{{}^3\kern-.14em {\rm S}_1}
\def\diii{{}^3\kern-.14em {\rm D}_1}

\newcommand{\Lcsb}{\Lambda_{\chi {\rm SB}}}

\def\si{{}^1\kern-.11em {\rm S}_0}
\def\siii{{}^3\kern-.11em {\rm S}_1}
\def\diii{{}^3\kern-.11em {\rm D}_1}
\def\pislash{ {\pi\hskip-0.6em /} }

\def\nopi{ {\rm EFT}(\pislash) }

\def\lsim{\mathrel{\rlap{\lower4pt\hbox{\hskip1pt$\sim$}}
    \raise1pt\hbox{$<$}}}         
\def\gsim{\mathrel {\rlap{\lower4pt\hbox{\hskip1pt$\sim$}}
    \raise1pt\hbox{$>$}}}         

\def\pone{{}^3\kern-.14em P_1}
\def\pzero{{}^3\kern-.14em P_0}
\def\ptwo{{}^3\kern-.14em P_2}
\def\pislash{ {\pi\hskip-0.6em /} }
\def\nopi{ {\rm EFT}(\pislash) }

\newtheorem{exercise}{Exercise}

\documentclass{czjphys}         
\usepackage{epsfig}
\begin{document}
\title{Building light nuclei from neutrons, protons, and pions}
%
\authori{Daniel Phillips}
\addressi{Department of Physics and Astronomy, Ohio University,
Athens, OH 45701, USA}
\authorii{}     \addressii{}
\authoriii{}    \addressiii{}
\authoriv{}     \addressiv{}
\authorv{}      \addressv{}
\authorvi{}     \addressvi{}
%
\headauthor{Daniel Phillips}            
\headtitle{Building light nuclei}       
\lastevenhead{Building light nuclei} 
\pacs{21.45.+v, 21.30.Cb, 25.10.+s, 11.10.Gh}
\keywords{effective field theory, few-body systems, light nuclei,
electromagnetic reactions}
\refnum{A}
\daterec{XXX}    
\issuenumber{0}  \year{2001}
\setcounter{page}{1}
\maketitle
\begin{abstract}
In these lectures I first explain, in a rather basic fashion, the
construction of effective field theories. I then discuss some recent
developments in the application of such theories to two- and three-nucleon
systems.
\end{abstract}

\section{Introduction: what is an effective theory?}

An effective theory is a {\it systematic approximation} to some
underlying dynamics (which may be known or unknown) that is valid
in some {\bf specified regime}. An effective theory is not a
model, since its systematic character means that, in principle,
predictions of arbitrary accuracy may be made. However, if this is
to be true then a small parameter, such as the $\alpha$ of quantum
electrodynamics, must govern the systematic approximation scheme.
As we shall see here, in many modern effective theories the
expansion parameter is a ratio of two physical scales. For
instance, in effective theories of supersymmetric physics ``beyond
the standard model'' the ratio would be of $p$ or $m$, the
momentum or mass of a standard model particle, to $M_{\rm SUSY}$.
In an effective theory for finite-proton-size effects in the
hydrogen atom the small parameter would be $r_{\rm p}$, the proton
size, divided by $r_{\rm b}$, the Bohr radius. The smallness of
this parameter is then indicative of the domain of validity of the
effective theory (ET). In this sense effective theories, like
revolutions, carry the seeds of their own destruction, since the
failure of the expansion to converge is a signal to the user that
he or she is pushing the theory beyond its limits. Within the
radius of convergence of the ET the ET ``works'' because of the
following fundamental tenet:
\begin{quotation}
Phenomena at low energies (or long wavelength) cannot probe details of
the high-energy (or short-distance) structure of particles.
\end{quotation}
I suspect that most physicists subscribe to this tenet: indeed, if the
tenet were not true, then physics (other than calculations using a
``theory of everything'') would be impossible.

In this first section I will begin by discussing the basic ideas
of effective theories using a few simple examples from
undergraduate physics. In this way we will move from classical
effective theories, to classical effective field theories (EFTs),
to quantum effective field theories.  Of course, to understand the
latter one must be able to compute quantum-field-theoretic loop
graphs, so this requires a little more education than the standard
undergraduate curriculum contains---at least the undergraduate
curriculum at American Universities! Nevertheless, I will attempt
to present the work at a level that is understandable by someone
who has completed a first course on quantum field theory, but has
not necessarily yet studied the theory of regularization and
renormalization.

In Section 2 I will begin to focus on the nucleon-nucleon (NN)
system. This will first necessitate some definitions of terms,
notation, and so forth. I then move on to discuss the special
issues stemming from the presence of shallow bound states in the
NN problem. After displaying one solution to this difficulty, I
will define and employ an effective field theory which takes into
account the presence of shallow bound states, but other than this
only contains neutrons and protons as dynamical degrees of
freedom. I will give examples of the success of this EFT, known as
$\nopi$, in computing (very-)low-energy electromagnetic
observables in the NN system.

The extension of this work to the three-body problem raises some
intriguing problems of renormalization. I will attempt to
elucidate these in Section 3, where I draw on the work of Bedaque,
Hammer, and van Kolck, to show how, after some thought and
interesting discoveries, $\nopi$ can be applied to the NNN
problem.

Finally, in Section 4 I give a brief tour of results in an
effective field theory with pions. The effective field theory of
QCD in which nucleons and pions are the degrees of freedom is
chiral perturbation theory ($\chi$PT). We were fortunate to have
one of the founders of $\chi$PT, and indeed a pioneer in the field
of EFTs, lecturing at this school. Prof.~Leutwyler's lectures in
this volume should be read in conjunction with the material
presented here. Indeed, this article should be regarded as little
more than light reading on the subject of nuclear EFTs.  It is
very far from being a thorough review on the topic. The reader who
wishes to study the subject in detail should consult the reviews
Ref.~\cite{vK99,Be00} which contain much more information than
does this manuscript. These two reviews also contain full
references to the original literature, a job I have not tackled in
any systematic way here.

\subsection{Some very simple effective theories}

\subsubsection{Gravity for $h < R$}
One of the simplest effective theories I know is one that is
learned by high-school physics students. It concerns the standard
formula for the gravitational potential-energy difference of an
object of mass $m$ which is raised a height $h$ above the Earth's
surface:

\begin{equation}
\Delta U=mgh,
\label{eq:mgh}
\end{equation}
where $g$ is the acceleration due to gravity

Of course from Newton's Law of Universal Gravitation (itself an
effective theory, valid in the limit of small space-time curvature),
we have
\begin{equation}
\Delta U=-{{GMm} \over {r_{\rm f}}} + {{GMm} \over {r_{\rm i}}},
\end{equation}
for an object whose distance from the Earth's centre is initially
$r_{\rm i}$ and finally $r_{\rm f}$. Now, if we write
\begin{equation}
r_{\rm f}=r_{\rm i} + h,
\end{equation}
and assume that $r_{\rm i} \approx R$, the radius of the Earth, then
\begin{equation}
\Delta U=m {{G M} \over {R^2}} {R \over {R + h}} h.
\end{equation}
Identifying $\frac{GM}{R^2}=g$ we see that Eq.~(\ref{eq:mgh}) is only
the first term in a series, which converges as long as $h/R < 1$:
\begin{equation}
\Delta U=m g h \left(1 - \frac{h}{R} + \frac{h^2}{R^2} + \ldots\right).
\label{eq:ETgrav}
\end{equation}
If we try to apply this theory to a satellite in geosynchronous
orbit ($h \gg R$) the series will not converge. But for the space
shuttle ($h \sim$ a few hundred km) this series should converge
fairly rapidly. Equation~(\ref{eq:mgh}) is the first term in the
effective theory expansion (\ref{eq:ETgrav}) for the gravitational
potential energy, with that ET being valid for $h < R$.

\subsubsection{Effective theories in the hydrogen atom}

Presumably, the hydrogen atom is ultimately described in terms of
string theory, or some other fundamental theory of physics.
Nevertheless, to very good precision, we can use quantum
electrodynamics (QED) as an effective theory to compute its
spectrum. The reason why we can ignore corrections to QED from
physics at the Planck scale when calculating the hydrogen-atom
spectrum will become clear below.

In the case of the hydrogen atom there is an effective theory for
QED that is valid up to corrections suppressed by one power of
$\alpha=e^2/(4 \pi)$, the fine-structure constant. That effective
theory is known as the Schr\"odinger equation with the Coulomb
potential. The radial wave function $u_{nl}(r)$ obeys the
differential equation~\footnote{Throughout I work in units where
$\hbar=c=1$.}:
5
\begin{equation}
-{1 \over {2 m_{\rm e}}} {{d^2 u_{nl}} \over {d r^2}} + {{l (l+1)} \over r^2}
u_{nl}(r) - {\alpha \over r} u_{nl}(r)=E_{n} u_{nl}(r).
\label{eq:Hatomde}
\end{equation}
The solution, for the lowest-energy eigenstate ($n=1$; $l=0$) is, of course:
\begin{equation}
u_{10}(r)={\cal N}\, \exp(-\alpha m_{\rm e} r)=
 {\cal N}\, \exp(-r/r_{\rm b}) \, ,
\end{equation}
where $r_{\rm b}=(\alpha m_{\rm e})^{-1} \approx (4 \mbox{ keV})^{-1} \approx
0.5~\AA$, and $\cal N$ is determined by the normalization condition.
The corresponding eigenvalue is
\begin{equation}
E_{10}=-{1 \over 2m_{\rm e}} \left({1 \over r_{\rm b}}\right)^2.
\end{equation}
The Bohr radius, $r_{\rm b}$, sets the scale for most phenomena
associated with the electron in the Hydrogen atom. In particular,
the typical momentum of the electron is $\sim 1/r_{\rm b}$, which
means that relativistic corrections to the Schr\"odinger equation
are suppressed by $(m_{\rm e} r_{\rm b})^{-2}=\alpha^2$, thereby
validating the non-relativistic treatment. Note that if we were
discussing muonic hydrogen the energy and distance scales would be
very different, since $r_{\rm b}^{\mu} \approx \frac{1}{200}
r_{\rm b}^{e}$.

The Bohr radius is large compared to the size of the proton,
$r_{\rm p}$, and also compared to the scale of internal structure
of the electron. One sense in which the electron has internal
structure is that it is dressed by virtual photons.  In fact, the
Lamb shift is, in fact, just such an electron-structure effect, so
``finite-electron-size'' effects must be considered if very
accurate results are desired. If we for the moment ignore the
electron's internal structure and consider only the internal
structure of the proton we would replace the Coulomb potential
$-\alpha/r$, by the potential generated by an extended proton:
\begin{equation}
V({\bf r})=-{e^2 \over 4 \pi} \int \, {{\rho({\bf r}')
{\rm d}^3r'}\over {|{\bf r} - {\bf r}'|}},
\label{eq:Coulomb}
\end{equation}
with $\rho({\bf r}')$ the local electric charge density of the
proton at the point ${\bf r}'$. Now we make a multipole expansion,
in order to obtain:
\begin{equation}
V({\bf r})=-{e^2 \over 4 \pi r} \sum_{n=0}^\infty
\left({r_{\rm p} \over r}\right)^n
\int \, {\rm d}^3 r' \,  \rho({\bf r}') P_n(\hat{r} \cdot \hat{r}')
\left({r' \over r_{\rm p}}\right)^n~~~\mbox{for $r > r_{\rm p}$,}
\label{eq:multipole}
\end{equation}
with $P_n$ the nth Legendre polynomial.  Here, $\rho(\bf{r}')$
only has support for $r' < r_{\rm p}$, and so the integrals are
all numbers of order one. Since the solution of the differential
equation (\ref{eq:Hatomde}) is mainly sensitive to $r \sim r_{\rm
b}$, the expansion parameter here is $r_{\rm p}/r_{\rm b} \sim
1~\AA/1~{\rm fm}$, and so this series converges rapidly, with it
entirely permissible to evaluate the corrections for the finite
size of the proton in perturbation theory.  Nevertheless, an
accurate computation requires inclusion of the term of order
$(r_{\rm p}/r_{\rm b})^2$ in this expansion.

In fact, this term of $O((r_{\rm p}/r_{\rm b})^2)$ is the first
correction due to finite-size effects in Eq.~(\ref{eq:Hatomde}).
This is easily seen from Eq.~(\ref{eq:multipole}), since the
coefficient of the term of $O(r_{\rm p}/r_{\rm b})$ is zero, as
long as the proton's charge distribution is even under parity.
Thus consideration of the scales in the problem alone would lead
us to grossly overestimate the magnitude of the finite-size
effect.  It is the combination of scales and symmetry that leads
to an accurate estimate of the magnitude of the effects neglected
by assuming that the proton is point-like in
Eq.~(\ref{eq:Hatomde}).  These two principles:
\begin{itemize}
\item a ratio of scales generating an expansion parameter,

\item symmetries constraining the types of corrections that can appear,
\end{itemize}
inform the construction of most effective theories.

\subsection{Building a (classical) effective theory: the
scattering of light from atoms}

Next I want to discuss an example of ET-construction which first
appeared in print in the effective field theory lecture notes of
Kaplan~\cite{Ka95}. These lecture notes are an excellent
introduction to EFT, and this example provides a great
demonstration of ET construction. Here I have reworked some of the
notation, but the basic idea is as in Ref.~\cite{Ka95}.

Consider the scattering of low-energy light from an atom.  The
question we must answer is: What is the Hamiltonian that describes
the interaction of the atom with the electromagnetic field of the
incoming light? To do this, we first consider the scales in the
problem: the energy of the electromagnetic field, $\omega$ is
assumed small compared to the spacing of the atomic levels,
$\Delta E$, and the inverse size of the atom. We will assume in
turn that all of these scales are much smaller than the mass of
the atom. Using $r_{\rm b}$ to estimate $\Delta E$ we have the
following hierarchy of scales:
\begin{equation}
\omega \ll \Delta E \sim {1 \over m_{\rm e} r_{\rm b}^2} \ll {1 \over r_{\rm b}} \ll
M_{\rm atom}.
\label{eq:hier}
\end{equation}

Meanwhile the symmetries of the theory will be electromagnetic
gauge invariance [$U(1)_{\rm em}$], rotational invariance, and
Hermitian conjugation/Time reversal. These symmetries constrain
the types of terms that we can write in our Hamiltonian. Firstly,
the constraint of gauge invariance suggests that it is wise to
construct $H_{\rm atom}$ out of the quantities ${\bf E}$ and ${\bf
B}$, rather than using the four-vector potential $A_\mu$. Then,
secondly, rotational invariance suggests that we employ quantities
such as ${\bf E}^2$ and ${\bf E} \cdot {\bf B}$ in $H_{\rm atom}$.
However, ${\bf E} \cdot {\bf B}$ is odd under time reversal, and
so we cannot write down a term proportional to it in $H_{\rm
atom}$. Meanwhile, terms such as ${\bf \nabla} \cdot {\bf B}$, and
${\nabla} \times {\bf E}$ may be included in $H_{\rm atom}$, but
then can be eliminated from the Hamiltonian using the field
equations for the electromagnetic field in the region around the
atom:
\begin{eqnarray}
\nabla \cdot {\bf E}=0; &\,& \nabla \cdot {\bf B}=0;\\
\nabla \times {\bf E}=0; &\,& \nabla \times {\bf B}=0.
\end{eqnarray}
This leaves us with:
\begin{equation}
H_{\rm atom}=a_1 {\bf B}^2 + a_2 {\bf E}^2
+ a_3 (\partial_0 {\bf B}^2)
+ a_4 (\partial_0 {\bf E}^2)
+ a_5 ({\bf E} \cdot {\bf B})^2 + \ldots
\end{equation}
In spite of our cleverness in constraining the terms that may
appear, we are still left with infinitely many operators that can
contribute to $H_{\rm atom}$. How are we to organize all of these
contributions?

\subsubsection*{Interlude: naive dimensional analysis}

The answer lies in the scale hierarchy established in
Eq.~(\ref{eq:hier}), together with a simple technique known as
naive dimensional analysis (NDA).  This works as follows:
consider, for instance, the operator ${\bf B}^2$.  Counting powers
of energy/momentum we see that it carries four powers of energy,
which we write as:
\begin{equation}
[{\bf B}^2]=4.
\end{equation}
However, $H_{\rm atom}$ must have dimensions of energy. It follows
that $a_1$ and $a_2$ must each carry three negative powers of energy/momentum:
\begin{equation}
[a_1]=[a_2]=-3,
\end{equation}
that is to say:
\begin{equation}
a_1,a_2={1 \over (\mbox{some energy scale})^3}.
\end{equation}
Now we ask what energy scales are present in the problem and so
might appear in the denominator here. The photon energy $\omega$
cannot appear in the denominator since the scale that occurs there
should refer to a property of the atom. Any of the scales
$r_0^{-1}$, $\Delta E$, or $M_{\rm atom}$ could be involved
though. The most conservative estimate would be that:
\begin{equation}
a_1,a_2 \sim {1 \over (\Delta E)^3}.
\end{equation}
However, very low-energy photons cannot probe the quantum level
structure of the atom: they should interact with the entire atom
in an essentially classical way. Thus, $\Delta E$ cannot occur in
the denominator of this lowest-dimensional term in the
Hamiltonian, and so we deduce that $a_1$ and $a_2$ must scale with
$r_0$, i.~e.:
\begin{equation}
a_1,a_2 \sim r_0^3,
\end{equation}
where the $\sim$ usually indicates that the coefficient here could
be a 3 or a 1/3 (or a -3 or a -1/3) but will generally be a number
of order one~\footnote{There is a subtlety here: since this is an
electromagnetic interaction the argument here suffices to get the
scaling with $\omega$ correct, but it does not count powers of
$\alpha_{\rm em}=1/137$ which is an additional small parameter in the
problem.}. $a_1$ and $a_2$ are in fact proportional to the
electric and magnetic polarizabilities of the atom~\cite{Hl00}.

Meanwhile, the operators multiplying the coefficients $a_3$ and
$a_4$ have dimension 5. Thus, $[a_3]=[a_4]=-4$, and so $a_3$ and
$a_4$ carry one more energy-scale downstairs as compared to $a_1$
and $a_2$. Conservatively, we assign the scaling:
\begin{equation}
a_3, a_4 \sim {r_0^3 \over \Delta E}.
\end{equation}
Similar estimates can be made for the other terms in $H_{\rm
atom}$. The key point is that since ${\bf B}^2$ and ${\bf E}^2$
are the lowest dimension operators allowed by the symmetries and
not already constrained by field equations, they will give the
dominant effect in $H_{\rm atom}$ for low-energy photons. Any
higher-order effects will be suppressed by at least $\omega/\Delta
E$, i.~e.:
\begin{equation}
H_{\rm atom}=r_0^3 \left[\tilde{a_1} {\bf B}^2 + \tilde{a_2} {\bf E}^2
+ O\left({\omega \over \Delta E}\right)\right],
\label{eq:Hatomnat}
\end{equation}
where $\tilde{a_1}$ and $\tilde{a_2}$ are now dimensionless
numbers. It is straightforward to turn this result into a
prediction for the photon-atom cross section. Since $[\sigma]=-2$
and the cross section results from squaring the quantum-mechanical
amplitude arising from the Hamiltonian (\ref{eq:Hatomnat}) we
discover that
\begin{equation}
\sigma \sim \omega^4 r_0^6 \[1 + O\({\omega \over \Delta E}\)\],
\label{eq:blue}
\end{equation}
where the $\sim$ disguises the hard work needed to figure out all
the factors of 2, $\pi$ and so forth that really go into deriving
$\sigma$!  The strong dependence of $\sigma$ on $\omega$ is, of
course, the reason the sky is blue---as pointed out in
Ref.~\cite{Ka95} or in Ref.~\cite{Jackson}, where the constant of
proportionality in Eq.~(\ref{eq:blue}) is worked out in detail!

\subsection{A classical effective {\it field} theory: Fermi electroweak theory}

Equation (\ref{eq:Hatomnat}) is an effective expression for the
classical Hamiltonian that is valid at long wavelength, or
equivalently, for low-energy electromagnetic fields. In general
effective field theories are derived for low energies, although
this need not be the case.

\begin{figure}[t]
\centerline{\epsfysize=1.2in \epsfbox{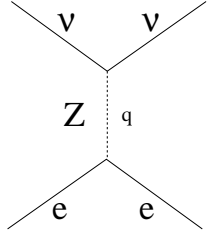}}
\vskip -5 mm
\caption{A Feynman diagram for neutrino-electron
scattering via exchange of a Z${}^0$ particle.}
\label{fig-Zexchange}
\end{figure}

A canonical example of a low-energy effective field theory is
Fermi's electroweak theory. This is an effective field theory that
can be used to compute, say, low-energy electron-neutrino
scattering. The only particles explicitly appearing in this theory
are electrons and neutrinos. By contrast, in the standard model,
the electrons and neutrinos interact by the exchange of W and Z
bosons. If we wish to compute the scattering of neutrinos from
electrons we could compute the full standard model amplitude for
diagrams such as Fig.~\ref{fig-Zexchange}~\cite{AH84}:
\begin{equation}
{\cal A}=\left({-{\rm i} g \over 2 \cos \theta_{\rm W}}\right)
\(\bar{\nu} P_{\rm L} \gamma_\mu \nu\) {{\rm i} \over q^2 - M_{\rm Z}^2}
\left({-{\rm i} g \over 2 \cos \theta_{\rm W}}\right) \(\bar{e} \gamma^\mu Q e\);
\label{eq:SMamplitude}
\end{equation}
with:
\begin{eqnarray}
P_{\rm L,R}&=&\frac{1}{2}(1 \mp \gamma_5);\\
Q&=&(-1 + 2 \sin^2 \theta_{\rm W}) P_{\rm L} + 2 \sin^2 \theta_{\rm W} P_{\rm R}.
\end{eqnarray}
and $q=p'-p$ being the change in momentum of the neutrino.
Relating $q^2$ to laboratory quantities, we see that:
\begin{equation}
q^2=-4 E_{\rm lab} E'_{\rm lab} \sin^2\({\theta_{\rm lab} \over 2}\),
\end{equation}
where $E_{\rm lab}$ ($E'_{\rm lab}$) and $\theta_{\rm lab}$ are the
initial (final) energy and scattering angle of the neutrino in the
lab. system. So, if $E_{\rm lab} \ll M_{\rm Z}$, then we can expand the
Z-propagator in Taylor series. The leading term in this series is
then:
\begin{equation}
{\cal A}={\rm i} \sqrt{2} G_{\rm F} \(\bar{\nu} P_{\rm L}
\gamma_\mu \nu\) \(\bar{e} \gamma^\mu Q e\),
\label{eq:FermiA}
\end{equation}
with:
\begin{equation}
G_{\rm F}={g^2 \over 4 \sqrt{2} \cos^2\theta_{\rm W}}
{1 \over M_{\rm Z}^2}={g^2 \over 4 \sqrt{2} M_{\rm W}^2}.
\end{equation}
The effective Lagrangian that generates the amplitude (\ref{eq:FermiA})
is:
\begin{equation}
{\cal L}_{\rm Fermi}=G_{\rm F} \sqrt{2} \(\bar{\nu}
\gamma_\mu P_{\rm L} \nu\) \(\bar{e} \gamma^\mu Q e\).
\label{eq:LFermi}
\end{equation}
This is the amplitude written down by Fermi many years ago for
beta-decay and associated processes. The argument given here for
deriving $G$ from the underlying theory goes under the rubric
``tree-level matching''.

Note that while the range of the neutrino-electron interaction
(\ref{eq:SMamplitude}) is of order $1/M_{\rm Z}$, the interaction
(\ref{eq:LFermi}) is of zero range. All four fields in the
Lagrangian (\ref{eq:LFermi}) are to be evaluated at the same
space-time point, $x$. This approximation is sensible provided
that the neutrino's energy is such that its Compton wavelength is
much longer than $1/M_{\rm Z}$. In this circumstance the standard model
Lagrangian may be approximated by a Lagrangian in which the
electrons and neutrinos interact only when they are right on top
of each other, since then the neutrino cannot resolve the finite
range of its interaction with the electron. A more accurate
approximation to the standard-model may be obtained by expanding
the Z-boson propagator in powers of $q^2/M_{\rm Z}^2$. These
higher-order pieces of the Fermi electroweak theory Lagrangian
will give rise to operators of a higher dimension than the
operator (\ref{eq:LFermi}), which has dimension:
\begin{equation}
2 [\nu] + 2 [e]=6.
\end{equation}

This is a crucial point, since the operator $\bar{\nu} \nu \bar{e} e$
is the lowest-dimension one we can construct out of the four lepton
fields. This means that even if we knew nothing about electroweak
interactions, we would still expect an interaction like
(\ref{eq:LFermi}) to be the leading term describing low-energy
electron-neutrino scattering.  Furthermore, since $[{\cal L}]=4$, and
$[\psi]=3/2$, we have $[G_{\rm F}]=-2$, and thus if we also had some inkling
that electroweak physics was determined by the one energy scale $M_{\rm Z}
\sim M_{\rm W} \sim 100$ GeV, we would expect:
\begin{equation}
G_{\rm F} \sim {1 \over ({\rm 100~GeV})^2}.
\label{eq:estimateG}
\end{equation}

Of course, this argument is ahistorical, since Fermi did now know
$M_{\rm Z}$ or $M_{\rm W}$. However, the argument can be turned around: if Fermi
had suspected that beta decay was due to the exchange of some heavy
particle he could have inferred its approximate mass from measurements
of $G_{\rm F}$.

\subsection{Effective {\it quantum} field theories: Scalar theory
in 5+1 dimensions}

An essential---perhaps the essential---feature of {\it quantum}
field theories is that one can compute loop corrections to the
tree-level result. Now in fact Fermi Electroweak Theory was used
in the previous section purely as a classical field theory of
Dirac particles, since only the tree-level amplitude
(\ref{eq:SMamplitude}) was discussed. We cannot be said to have
derived an effective {\it quantum} field theory until we have
explained how to compute loops. Here we encounter an apparent
difficulty, because the effective field theories we have written
down in the previous two subsections are ``non-renormalizable''
quantum field theories.  Interactions such as (\ref{eq:LFermi})
are highly singular at short distances, and this leads to
divergences when loop graphs are computed. Such divergences are
not an insurmountable difficulty: they occur in quantum
electrodynamics, which is rendered predictive by the use of
subtractions in the misbehaving integrals and the input of two
pieces of physical data ($m_{\rm e}$ and $e$). However, in
``non-renormalizable'' field theories such as Fermi electroweak
theory the calculation of higher-loop graphs leads to divergences
with more and more complicated structures.  So much so that at
each new order in the loop expansion additional pieces of physical
data are needed in order to render the theory finite. This appears
to be a bad situation, since it suggests that the theory has no
predictive power, as it requires infinitely many inputs before it
can make a prediction. This conclusion is, however, hasty. In this
section I will explain how to compute loop graphs in a
non-renormalizable effective field theory, and indicate why the
theory has predictive power in spite of its non-renormalizability.

This will be done using an example effective field theory which
has no real uses that I am aware of. Although perhaps this has
changed with the recent interest in dilatons, radions, and so on,
in higher-dimensional field theories. The field theory is that of
a massive but light scalar particle in 5+1 dimensions. We will
take the symmetries of the theory to be six-dimensional Lorentz
covariance and invariance under the discrete transformation $\phi
\rightarrow -\phi$. This leaves us with infinitely many
interaction terms we can write down in the Lagrange density:
\begin{equation}
{\cal L}={1 \over 2} (\partial_\mu \phi)(\partial^\mu \phi) - {1 \over
2} m^2 \phi^2 - {\lambda \over 4!} \phi^4 - m^2 {\tilde{\lambda} \over
4!} \phi^4 - {\xi \over 2! 2!} (\partial_\mu \phi)^2 \phi^2 - {\zeta
\over 6!}  \phi^6.
\label{eq:L6}
\end{equation}
Here $[{\cal L}]=6$, $[\phi]=2$, and so $[\lambda]=-2$, implying
that all the interaction terms here are non-renormalizable.
However, the statement $[\lambda]=-2$ also implies that
\begin{equation}
\lambda={a \over \Lambda^2},
\end{equation}
where $a$ is a number of order one, and we have assumed that this
field theory is an EFT for an underlying theory with one key heavy
scale, $\Lambda$. Likewise:
\begin{equation}
\tilde{\lambda}={\tilde{a} \over \Lambda^4}; \quad
\xi={b \over \Lambda^6}; \quad \zeta={c \over \Lambda^6},
\end{equation}
where we will assume that $\tilde{a}$, $b$, and $c$ are all of
order one, i.~e. they are ``natural''.

What would we then expect for the threshold amplitude for $\phi
\phi$ scattering on the basis of NDA? In general, $T_{\rm th}$
will be a function of $m$, the light scalar mass, and so, since
$[T_{\rm th}]=-2$, we expect $T_{\rm th}$ to vary in the following
way:
\begin{equation}
T_{\rm th}={1 \over M^2} + \aleph {m^2 \over M^4} + \ldots +
\mbox{possible terms proportional to $m^2 \log\left({m^2 \over
M^2}\right)$},
\label{eq:Tvarn}
\end{equation}
here $M$ is just a convenient mass scale determined by the size of
the threshold amplitude at $m=0$, and $\aleph$ is order one.

\begin{figure}[t]
\centerline{\epsfysize=1.2in \epsfbox{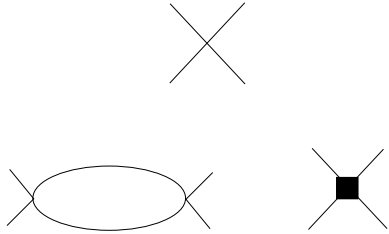}}
\vskip -5 mm
\caption{Feynman diagrams for $\phi \phi$ scattering at threshold in
six-dimensional $\phi^4$ theory. The first line represents the
leading-order result, from the vertex $\lambda \phi^4$, while the
second line includes the one-loop graph constructed from these
leading-order vertices, as well as the next-to-leading order vertex
$m^2 \tilde{\lambda} \phi^4$.}
\label{fig-6dphi4}
\end{figure}

By contrast, direct calculation with (\ref{eq:L6}) shows that the
``leading-order'' (LO) contribution to $T_{\rm th}$, coming from
the graph on the first line of Fig.~\ref{fig-6dphi4} is just:
\begin{equation}
T_{\rm th}^{\rm LO}={a \over \Lambda^2}.
\end{equation}
But leading-order in what? In fact, it will prove to be leading
order in an expansion in $m/\Lambda$. And we already see that if
the scale $M$ is of order $\Lambda$, which it should be for a
generic quantum mechanics problem, then adjusting $a$ to reproduce
the first term in Eq.~(\ref{eq:Tvarn}) will only require that $a$
be of order one. Systems where $M$ is not of order
$\Lambda$---such as the NN system---are referred to as
``fine-tuned'' and require special treatment. We will discuss this
in the next section.

The next order in the $m/\Lambda$ expansion is provided by the two
graphs on the second line of Fig.~\ref{fig-6dphi4}. Both of these
graphs are $O(m^2/\Lambda^4)$. However, the loop graph has a
logarithmic divergence multiplying $m^2/\Lambda^4$. Using
dimensional regularization to evaluate the loop integral in $6 -
\epsilon$ dimensions we find that the overall contribution of the
two graphs to $T_{\rm th}$ is:
\begin{equation}
T_{\rm th}^{\rm NLO}=-{\lambda^2 \over 2}{1 \over (4 \pi)^{3}}
\left(4 \pi \mu^2\right)^{\epsilon/2}
\Gamma\left({\epsilon \over 2} - 1\right) (m^2)^{1 - \epsilon/2}
+ m^2 \tilde{\lambda}.
\end{equation}
(If you are unfamiliar with dimensional regularization you should
consult a field theory book such as Ref.~\cite{Ryder} for a primer
on this topic.)  Making a Laurent expansion of the Gamma function
for small $\epsilon$, and expanding the other functions present in
powers of $\epsilon$ too, we find that as $\epsilon \rightarrow
0$:
\begin{equation}
T_{\rm th}^{\rm NLO}=m^2\left\{{\lambda^2 \over 2} {1 \over (4 \pi)^3}
\left[{2 \over \epsilon} + c + \log\left(\mu^2 \over m^2\right)
\right] + \tilde{\lambda}\right\},
\label{eq:TthNLO}
\end{equation}
where $c$ is an $m$-independent constant, whose computation
is left as an exercise:
\begin{exercise}
Show that $c=1 - \gamma + \log(4 \pi)$, where $\gamma$ is the
Euler-Mascharoni constant.
\end{exercise}

Putting in our NDA expectations $\tilde{\lambda}=a_2/\Lambda^4$
and $\lambda=a/\Lambda^2$ we see that $T_{\rm th}^{\rm NLO}$ is
indeed of order $m^2/\Lambda^4$ and so will be suppressed relative
to $T_{\rm th}^{\rm LO}$, as long as the scalar is light relative
to $\Lambda$.  However, it appears that $T_{\rm th}^{\rm NLO}$ is
a small infinity, since part of the expression for it diverges as
$\epsilon \rightarrow 0$!  In order to resolve this dilemma we
need a piece of physical data in order to fix the counterterm
$\tilde{\lambda}$. In just the same way, in QED the bare electron
mass is chosen to be infinite, with that infinity used to cancel
off infinite contributions from the one-loop self-energy, and a
physical parameter, namely $m_{\rm e}$, must be specified in order
to render the theory predictive.

In our case we will suppose that we know the coefficient $\aleph$
in Eq.~(\ref{eq:Tvarn}), from calculations in the underlying
theory, say:
\begin{equation}
\left.{\partial T_{\rm th} \over \partial m^2}\right|_{m^2=m_0^2}
=T_0'.
\end{equation}
In this case $\tilde{\lambda}$ must be chosen so that:
\begin{equation}
T_0'=\tilde{\lambda} + {\lambda^2 \over 2} {1 \over (4 \pi)^3}
\left[{2 \over \epsilon} + c + \log\left(\mu^2 \over m_0^2\right)\right].
\label{eq:renorm}
\end{equation}
Note that the constant $\tilde{\lambda}$ is renormalization-scheme
dependent: choices of ``minimal subtraction'' (MS) or ``modified
minimal subtraction'' ($\overline{\rm MS}$), will result in
different definitions for $\tilde{\lambda}$.  However, the
physical prediction for $T_{\rm th}$ is independent of
renormalization scheme: if we invert (\ref{eq:renorm}) and use it
to remove $\tilde{\lambda}$ from (\ref{eq:TthNLO}) then we find
that the infinity as $\epsilon \rightarrow 0$ has been absorbed in
$\tilde{\lambda}$, and:
\begin{equation}
T_{\rm th}^{\rm NLO}=m^2 \left[T_0' + {\lambda^2 \over 2}{1 \over (4 \pi)^3}
\log\left({m_0^2 \over m^2} \right)\right],
\end{equation}

What we see here is that the coefficient of the logarithmic term
that describes the non-analytic variation of $T_{\rm th}^{\rm
NLO}$ with $m$ is {\it predictive}, and is calculable
order-by-order in the low-energy theory. This non-analytic
behaviour arises from the long-distance, low-momentum, infra-red
physics of the light-particle loops. The short-distance physics is
encoded in the coefficient $\tilde{\lambda}$, or, equivalently, in
the piece of physical data $T_0'$.

The example we have examined here is really quite simple, but the
behaviour it displays is generic. If we consider {\it any} loop
graph contributing to the process $\phi \phi \rightarrow \phi
\phi$ then its superficial degree of divergence is:
\begin{equation}
d=\sum_i n_i d_i + 2 L,
\label{eq:ned}
\end{equation}
where $n_i$ is the number of vertices of type $i$, $d_i$ is the
number of powers of momentum that this type of vertex carries and
$L$ is the number of loops in the graph. However, if dimensional
regularization and a mass-independent subtraction scheme are
employed the sum of this loop and appropriate counterterms must
carry $d$ powers of $m$ and/or momentum $p$. Hence it must behave
like:
\begin{equation}
{(p,m)^d \over \Lambda^{d+2}}
f\left({p \over \Lambda},{m \over \Lambda}\right),
\end{equation}
where $f$ is a dimensionless function of $p$ and $m$.  This has
the further consequence that in order to ensure that all of the
appropriate counterterms are present, and so all divergences can
be removed, we must include {\it all} possible local counterterms
in ${\cal L}$ which scale as $(p,m)^d$.  In other words, we must
include in our original Lagrangian all possible terms of that
order which are consistent with the symmetries of the theory.
Ultimately this amounts to calculating the loop graph in order to
obtain the function $f$ and then making as many subtractions as
are necessary to render $f$ finite. Thus the effective field
theory, although non-renormalizable, can make predictions for the
mass and energy dependence of observables.

A key point here is that in any non-renormalizable theory the
degree of divergence $d$ of graphs increases as more vertices and
more loops are added. This had been regarded as a ``bug'' in
non-renormalizable field theories, but now we see it is instead a
``feature''~\footnote{Credit to Tom Cohen for this remark.}.
Non-renormalizability becomes a ``feature'' because it means that
graphs with more loops and/or more complicated vertices carry more
powers of the small scales $(p,m_\pi)$, and so are suppressed
relative to tree-level and low-loop-order graphs. This fascinating
paradigm shift is discussed further in
Refs.~\cite{Brownbook,Le90,Le97}.

In a non-renormalizable field theory computing all loop graphs and
tree-level contributions that carry up to $d$ powers of $p$ and
$m_\pi$ we will obtain a prediction for the $\phi \phi \rightarrow
\phi \phi$ amplitude which is accurate up to terms which are
suppressed by an amount $p^2/\Lambda^2$ and $m^2/\Lambda^2$
relative to those that are included. EFT predictions such as the
one obtained above for the mass dependence of the scalar-scalar
scattering amplitude in six dimensions can thus be systematically
improved by computing more loops, including more counterterms etc.

\subsection{Summary: the EFT algorithm}

Thus we can develop what I shall refer to as ``The EFT
algorithm''. The steps in this procedure, which can be used to
build a low-energy theory of just about anything, are (with
apologies to the experts who will recognize that this is somewhat
simplified) as follows:
\begin{enumerate}
\item Identify the degrees of freedom of interest.

\item Identify the low-energy scales and high-energy scales in the theory.
The ratios will form expansion parameters, and so a good scale separation
will facilitate construction of the EFT.

\item Identify the symmetries of the low-energy theory.

\item Choose the accuracy required. This, together with the size of the
largest expansion parameter, say $x$, will determine the order $n$ to
which observables must be computed.

\item Write down all possible local operators, consistent with
the symmetries, which have dimensions up to that order:
\begin{equation}
{\cal L}={\cal L}_D + {\cal L}_{D+1} + {\cal L}_{D+2}
+ \ldots + {\cal L}_{D + n}.
\end{equation}

\item Derive the behaviour of loops, that is to say, the analog of
Eq.~(\ref{eq:ned}) for this particular theory. This will tell you
how many loops you need to calculate to have a prediction that
is accurate to $n$th order in $x$.

\item Calculate the loops and renormalize them.
\end{enumerate}

At this point you should have a prediction which will be accurate
up to terms whose relative size will be set by $x$ (as long as the
mass scales which generated $x$ can actually produce a term at the
next order). Now, in fact, to check that the theory is really
converging in the expected manner an $n+1$st order calculation is
necessary. Comparison with experimental data can also be made, but
note that in addition to the usual experimental errors EFTs carry
an intrinsic theoretical error, of size $x^{n+1}$. This error is
an honest reflection of the impact of our ignorance of high-energy
physics on the observable in question.

\section{The NN system in $\nopi$}

In this section I will implement this ``EFT algorithm'' for
nucleon-nucleon scattering at low energies, i.~e.  scattering
energies much lower than $m_\pi^2/M$. At such energies the Compton
wavelength of the nucleons is such that even the longest-ranged
part of the NN force, that due to one-pion exchange, appears
short-ranged, and so we can approximate the NN interaction by a
potential of zero range.  As we shall see this is an old story,
dating back at least 50 years, and some of the recent work on EFT
at these energies involved rediscovering the wisdom of the
ancients. However, EFT provides new wrinkles to the old tales, and
I will argue that these make the techniques developed by Hans
Bethe and others all those years ago even more useful.

\subsection{Review of some basic NN facts}

To begin with, and to establish notation let us review some basic
NN scattering facts. If the NN interaction $V$ falls off
sufficiently quickly at large $r$ then the asymptotic wave
function has the form:
\begin{equation}
\psi({\bf r})= \exp({\rm i} {\bf k}_{\rm i} \cdot {\bf r})
+ f(k,\theta) {\exp({\rm i} k_{\rm f} r) \over r},
\end{equation}
where $k_{\rm i}^2=k_{\rm f}^2 \equiv k^2=ME$ with $E$ the energy
in the centre-of-mass system, and $\cos \theta=\hat{k}_{\rm i}
\cdot \hat{k}_{\rm f}$. The differential cross section for NN
scattering is then given by
\begin{equation}
{d \sigma \over d \Omega}=|f(k,\theta)|^2.
\end{equation}
Of course, if $V$ is a central potential one may make a
partial-wave expansion for $f$. In the absence of spin this has
the form:
\begin{equation}
f(k,\theta)={1 \over k}\sum_l (2 l + 1) \exp({\rm i} \delta_l)\,
\sin \delta_l\, P_l(\cos \theta) \, .
\end{equation}
At this point the reader might reasonably complain that the NN
problem involves spin {\it and} a non-central potential, but let
me leave these complications aside for the moment, because for the
${}^1{\rm S}_0$ channel, which I will deal with first, neither actually
matter. In this, or in any S-wave channel, the asymptotic radial
wave function $u$ has the form:
\begin{equation}
u_0(r) \stackrel{r \rightarrow \infty} \longrightarrow v_0(r) \sim
\sin(k r + \delta_0),
\end{equation}
with $\delta_0$ the S-wave phase shift. The S-matrix itself is
$S=\exp(2{\rm i} \delta_0)$ and this leads to a T-matrix (in the
normalization used here):
\begin{equation}
T=-{4 \pi \over M} {1\over {k \cot\delta - {\rm i} k}}.
\end{equation}

This expression for the T-matrix incorporates the constraint of
unitarity, i.~e. probability conservation. An expansion for $T$,
such as that obtained in EFTs like the one discussed above, only
incorporates unitarity order-by-order in the EFT expansion. There
is, however, a well-known expansion for the amplitude $T$ which
obeys (two-body) unitarity exactly. That is the effective range
expansion (ERE), which is a power-series expansion for $k \cot
\delta$. The ERE is a power series expansion for the real part of
$1/T$. A simple derivation, valid for any energy-independent
potential, shows that:
\begin{equation}
k \cot \delta(k) - \lim_{k \rightarrow 0} k \cot \delta(k)={1 \over 2}
k^2 \rho(E,0),
\label{eq:kcotd}
\end{equation}
where:
\begin{equation}
{1 \over 2} \rho(E_1,E_2)=\int_0^\infty \left[v_{E_1}(r)
v_{E_2}(r) - u_{E_1}(r) u_{E_2}(r)\right] dr,
\label{eq:rho}
\end{equation}
with $u_E(r)$ the S-wave radial wave function for the potential $V$ at
energy $E$, while
\begin{equation}
v_E(r)={\sin(k r + \delta(k)) \over \sin \delta(k)},
\end{equation}
is a wave function that agrees with $u$ in the asymptotic regime
but differs from it inside the range of the potential, $R$.

\begin{exercise}
Derive Eq.~(\ref{eq:kcotd}) by considering an energy-independent
potential and analyzing the Schr\"odinger equation at two different
energies $E_1$ and $E_2$.
\end{exercise}

Thus, the region of support for the integration in
Eq.~(\ref{eq:rho}) is restricted to $r \leq R$. This suggests that
we may make a Taylor expansion for $\rho$ in powers of $E$ and
that such an expansion should have a radius of convergence $\sim
1/R$:
5
\begin{equation}
\rho(E,0)=\rho(0,0) + E {\partial \rho(E,0) \over \partial E} + \ldots.
\label{eq:rhoexp}
\end{equation}
Combining Eqs.~(\ref{eq:kcotd}) and (\ref{eq:rhoexp}) yields Bethe's
effective-range expansion~\cite{Be49,BL50}:
\begin{equation}
k \cot \delta(k)=-{1 \over a} + {1 \over 2} r_0 k^2 - P r_0^3 k^4 +
\ldots.
\end{equation}
This can be converted to an expansion for $\delta$ itself using
the power series for the arccot, but the result only has a radius
of convergence $\sim 1/a$. Thus, if $R \ll a$ the effective range
expansion is valid to much higher energies than a simple
power-series expansion for $\delta$. The effective range expansion
was developed to treat precisely such a problem: NN scattering at
low energies.

\begin{exercise}
Show that for an energy-independent Hermitian potential of range $R$
the effective-range is bounded from above by~\cite{PC97}:
\begin{equation}
r_0 \leq 2 \left(R - {R^2 \over a} + {R^3 \over 3 a^2}\right).
\end{equation}
\end{exercise}

\begin{figure}[t]
\centerline{\epsfysize=2.5in \epsfbox{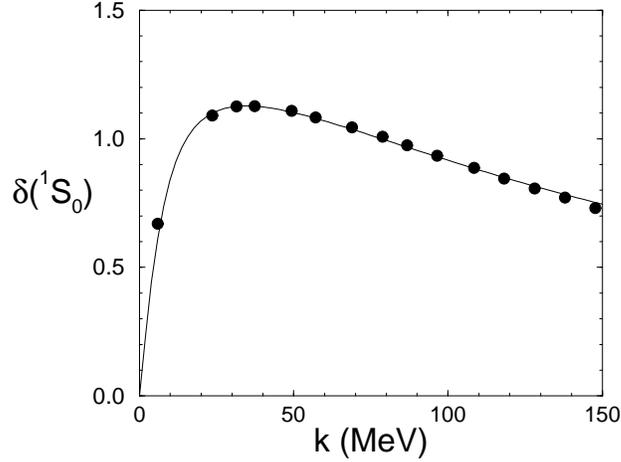}}
\vskip -8mm
\caption{The effective-range expansion in the ${}^1{\rm S}_0$ channel (solid
line) as compared to the Nijmegen phase-shift analysis (dots)~\cite{St93}.}
\label{fig-1S0ERE}
\end{figure}

In the ${}^1{\rm S}_0$ NN channel the central values of the
effective-range parameters~\footnote{In fact the shape parameter
$P$ is not well-determined by the experimental data, but it is
definitely small.} are~\cite{Ka98B}:
\begin{eqnarray}
&&a\left({}^1{\rm S}_0~{\rm np}\right)=-23.71~{\rm fm};\nonumber\\
&&r_0\left({}^1{\rm S}_0~{\rm np}\right)=2.73~{\rm fm};\nonumber\\
&&P\left({}^1{\rm S}_0~{\rm np}\right)=0.06~{\rm fm},
\end{eqnarray}
while in the ${}^3{\rm S}_1$ channel we have~\cite{dS95}:
\begin{eqnarray}
&&a\left({}^3{\rm S}_1\right)=5.42~{\rm fm};\nonumber\\
&&r_0\left({}^3{\rm S}_1\right)=1.75~{\rm fm}.
\label{eq:siiiexpt}
\end{eqnarray}
The leading-order ERE amplitude is then:
\begin{equation}
T_{\rm NN}^{{\rm ERE} (0)}={4 \pi a \over M} {1 \over 1 + {\rm i} a k},
\label{eq:TERE0}
\end{equation}
Meanwhile the second-order ERE amplitude is:
\begin{equation}
T_{\rm NN}^{{\rm ERE} (2)}=-{4 \pi \over M}
{1 \over -\frac{1}{a} + \frac{1}{2} r_0 k^2 - {\rm i} k}.
\label{eq:TERE2}
\end{equation}
In the ${}^1{\rm S}_0$ channel the amplitude (\ref{eq:TERE2})
reproduces the experimental data up to momenta $k \sim m_\pi$ (see
Fig.~\ref{fig-1S0ERE}), even though strictly speaking the radius
of convergence of the effective-range expansion is $m_\pi/2$.

The amplitude (\ref{eq:TERE0}) has a pole at momentum $k={\rm
i}/a$.  In the ${}^3{\rm S}_1$ system this pole corresponds to a
low-energy bound state. This analytic structure of $T(k)$ must be
reproduced if our effective theory is going to give a good account
of NN scattering data in the low-energy regime.

\begin{exercise}
Using the experimental numbers for the ${}^3{\rm S}_1$ channel from
Eq.~(\ref{eq:siiiexpt}) compute the energy of the deuteron pole as
given by the effective-range expansion. How does it compare with the
experimental number:
\begin{equation}
B=2.224575(9)~\rm{MeV}?
\end{equation}
\end{exercise}

\subsection{Low-energy poles in $\nopi$}

We now try to build an effective field theory to reproduce
(\ref{eq:TERE0}). Once this is achieved we will look at how to
reproduce the NN amplitude if terms of higher-order in the
effective-range expansion are incorporated, but we start with this
very simple problem. The only degrees of freedom in our EFT will
be the nucleons themselves, so we call this theory $\nopi$. The
scale hierarchy in the theory is:
\begin{equation}
k \ll m_\pi,
\end{equation}
where $k$ is the nucleon momentum and $1/m_\pi$, which from now on
will be denoted $1/\Lambda$, is the range of the force in the
``underlying'' theory. Note that here we have not initially
specified whether the scales $1/a$ and $1/r_0$ are low-energy
scales (like $p$) or high-energy scales (like $\Lambda$), although
since $1/a \approx 8$ MeV in the $\si$ channel, it better
ultimately be a low-energy scale!  The symmetries of the theory
will be translational invariance, rotational invariance (which
may, however, be broken due to operators from one-pion exchange),
and the Lorentz group for small Lorentz boosts. Only small boosts
are allowed, since the theory is only valid for energies such that
the nucleon momenta are much smaller than $m_\pi$, and so the
nucleons are non-relativistic.  Relativistic corrections can be
incorporated systematically by demanding that the theory is
Lorentz covariant order-by-order in $p/M$. (We shall not discuss
this further here, for details see Ref.~\cite{Ch99}.)  With this
in mind the Lagrangian for $\nopi$ is:
\begin{equation}
{\cal L}=N^\dagger \left({\rm i} D_0 + {{\bf D}^2 \over 2 M} \right) N +
{\cal L}_{\rm I}.
\label{eq:nopi1}
\end{equation}
The interaction Lagrangian, ${\cal L}_{\rm I}$ in principle contains all
possible $n$-body interactions of nucleons, organized as a series in
the dimensionality of the local operators that could contribute. For
the two-nucleon system the lowest-dimension operator in ${\cal L}_{\rm I}$
is $(N^\dagger N)^2$, which has dimension six. Increasing numbers of
derivatives and higher-body interactions can be included, but the
corresponding operators are of higher dimension. Thus first we examine
what we expect to be the ``leading-order'' (LO) theory: the NN
scattering amplitude which arises from a Lagrange density with only a
local momentum-independent four-nucleon operator.  Therefore, the
Lagrangian which we will use in our attempt to recover the amplitude
(\ref{eq:TERE0}) is:
\begin{equation}
{\cal L}_{\rm I} =  \ - \ \left(\mu \over 2\right)^{4-n} C_0
\left(N^{\rm T} P N\right)^\dagger\left(N^{\rm T} P N\right),
\label{eq:invar}
\end{equation}
where $P$ is a spin and isospin projector that projects the NN
pair onto the $LSJT$ combination that is of interest to us---in
this case, either the ${}^1{\rm S}_0$ or the ${}^3{\rm S}_1$, and
$n$ is the number of space-time dimensions. The factor
$(\mu/2)^{4-n}$ is included so that $[C_0]=-2$ even when the
theory is continued to fewer than four dimensions.

Now, naive-dimensional analysis as formulated in the previous
section implies that ${\bf D}^2$, as a dimension five operator, is
suppressed relative to the dimension-four operator $D_0$.
However, in a non-relativistic system the kinetic energy of our
system is the same order as its total energy, and so $D^2/M$ is of
the same order as $D_0$.  Furthermore, in a non-relativistic bound
(or quasi-bound) state such as we are considering here we also
know that both of these quantities are of the same order as the
potential energy. In this theory the NN potential is very simple.
It is given by:
\begin{equation}
\langle {\bf p}'|V|{\bf p} \rangle=C_0,
\label{eq:C0pot}
\end{equation}
with ${\bf p}$ (${\bf p}')$ the initial (final) relative momentum
of the NN system. If we want $V$ to be of the same order as the NN
kinetic energy then the $C_0$ operator cannot be treated in the
way suggested by naive dimensional analysis, since the naive
dimensional analysis estimate is:
\begin{equation}
C_0 \sim {1 \over M \Lambda}.
\label{eq:NDA}
\end{equation}
We would expect then, that $C_0$, as a dimension-six operator,
will be even more suppressed than ${\bf D}^2$. (The factor of $M$
appears here since this is a non-relativistic effective field
theory~\cite{LM97}.)

\begin{figure}[t]
\centerline{\epsfysize=1.0in \epsfbox{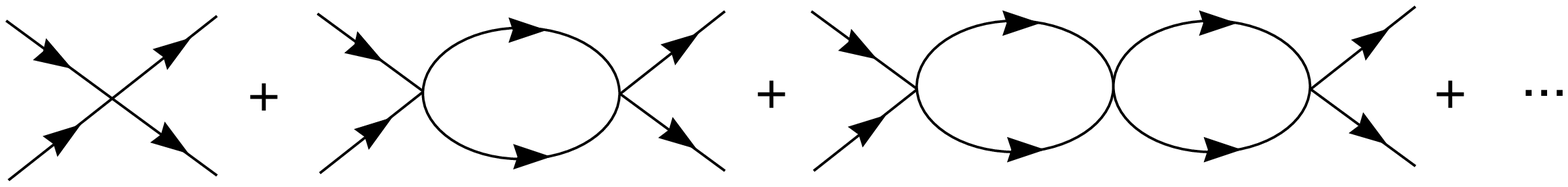}}
\vskip -8 mm
\caption{Feynman diagrams contributing to the
scattering amplitude from the Lagrange density in
Eqs.~(\ref{eq:nopi1}) and (\ref{eq:invar}). Figure courtesy M.~Savage.}
\label{fig-bubbles}
\end{figure}

This makes us wary of treating any contribution in the Lagrangian
(\ref{eq:invar}) in perturbation theory. But we will now see that
the Lagrange density (\ref{eq:invar}) is simple enough that the
sum of all Feynman diagrams contributing to NN scattering, which
is shown in Fig.~\ref{fig-bubbles}, can be found exactly.

Adding one additional loop to a Feynman graph adds one extra loop
integration, one extra non-relativistic Green's function, and one
extra $C_0$ vertex. Thus if $k$ were the ``typical'' momentum
scale of the problem, say the incoming NN relative momentum, then
the higher-order Born series graph would indeed be higher order.
It would contain an additional factor of
\begin{equation}
k^3 \times {M \over k^2} \times C_0
\end{equation}
as compared to its cousin which is one order lower in the Born
series for the potential (\ref{eq:C0pot}). Putting in the NDA
estimate (\ref{eq:NDA}) for $C_0$ indicates that adding a loop
should reduce the size of the diagram by one factor of
$k/\Lambda$.  This is another way to see what we have guessed in
the previous paragraph: if $C_0$ does scale according to
Eq.~(\ref{eq:NDA}) then there cannot be a low-energy NN bound
state, since the potential (\ref{eq:C0pot}) is ``small'' compared
to the kinetic energy, and so the Born series can be used to
calculate observables in perturbation theory.

If, however, we blithely go ahead and sum the Born series depicted
in Fig.~\ref{fig-bubbles} we obtain a $T$-matrix
\begin{eqnarray}
T_{\rm NN} & = &
\left[\
C_0 +  C_0 I_0 C_0 + C_0 I_0 C_0 I_0 C_0 + \ldots\ \right]
\ =\ {1 \over 1/C_0-I_0}
\ \ ,
\label{eq:geom}
\end{eqnarray}
where $I_0$ is the momentum-independent single-loop integral,
which in the NN center-of-mass frame has the form:
\begin{eqnarray}
I_0 & = &
-{\rm i}\left({\mu \over 2}\right)^{4-n}
\int {{\rm d}^n q\over (2\pi)^n} \
\left({{\rm i} \over E + q_0 -{\bf q}^2/ 2M + {\rm i}\eta}\right)
\left({{\rm i} \over - q_0 -{\bf q}^2/2M + {\rm i}\eta} \right)
\nonumber\\
& = & \left({\mu \over 2}\right)^{4-n}
\int {{{\rm d}}^{(n-1)}{\bf  q}\over (2\pi)^{(n-1)}}\
\left({1\over E  -{\bf q}^2/M + {\rm i}\eta}\right)
\label{eq:I0}
\end{eqnarray}
Here $|{\bf k}|=\sqrt{ME}$ the magnitude of the three-momentum of
each nucleon in the center-of-mass frame. Note that this integral
$I_0$ is linearly divergent, and so we have evaluated it in $n$
space-time dimensions.

In order to proceed further a regularization and renormalization
scheme must be specified. It is clear from matching
Eqs.~(\ref{eq:geom}) and (\ref{eq:TERE0}) to one another that the
value of $C_0$ will depend on both the value of $1/a$ and the
meaning assigned to the divergent part of $I_0$ by our
regularization and subtraction procedures.  No matter how we
define $I_0$, and hence the loop contributions, we can choose
$C_0$ so that (\ref{eq:TERE0}) is obeyed, but different
regularization and renormalization procedures amount to a
re-shuffling between contributions from the $C_0$ vertices and the
UV part of the loop integration (\ref{eq:I0}).

We begin by applying dimensional regularization with the MS
subtraction scheme. Evaluating the integral in Eq.~(\ref{eq:I0})
in the usual fashion of dimensionally-regulated loop integrals we
see that
\begin{equation}
I_0=-M\
(-|{\bf k}|^2- {\rm i}\epsilon)^{(n-3)/2}\, \left({\mu \over 2}\right)^{4-n}\,
\Gamma\left({3-n \over 2}\right)\, (4\pi)^{(1-n)/2} \, ,
\label{eq:loopi}
\end{equation}
The integral has been regulated by the continuation to $n$
dimensions. The linear divergence now manifests itself as a pole
of the $\Gamma$ function, which appears at $n=3$. However, the
vagaries of the $\Gamma$ function mean that the $n \rightarrow 4$
limit of the expression in Eq.~(\ref{eq:loopi}) is finite, so
straightforward evaluation just gives:
\begin{equation}
I_0^{\rm MS}=- \left({M\over 4\pi}\right)\, {\rm i}|{\bf k}|,
\label{eq:I0MS}
\end{equation}
and thus the linear divergence in the integral never appears in
the result for $I_0$. This is in contrast to the example in the
previous section, where a logarithmic divergence in the integral
in question revealed itself as a pole at the point $n=4$
($\epsilon=0$) once the integral was continued to a different
number of space-time dimensions.

Equation (\ref{eq:I0MS}) is the MS value for $I_0$. Re-examining
Eq.~(\ref{eq:geom}) we see that in this subtraction scheme each
successive term in the bubble-sum involves the product $C_0 M
|{\bf k}|$. Thus, in order to produce a pole in the amplitude at
momenta of order $1/a$ it is necessary that the coefficient $C_0$
scale as $C_0 \sim a/M $. In the ${}^3{\rm S}_1$ channel this
makes $C_0$ significantly larger than its NDA value, and in the
$\si$ channel this makes $C_0$ absolutely huge. This unnatural
scaling in MS, and the problems it leads to, were first discussed
by Kaplan, Savage, and Wise~\cite{Ka96}.  $C_0$ is simply
unnaturally large: in fact it is ill-defined for the case of a
threshold bound state, since $a \rightarrow \infty$ there.

The real solution to this difficulty lies in the development of a
new power counting~\cite{Lu95,vK97,Ka98B,Ka98A,vK98,Bi99}.
Subsequently it was shown how to implement this power counting on
a diagram-by-diagram basis, using a general subtraction procedure
called Power Divergence Subtraction~\cite{Ka98B,Ka98A} (PDS). In
PDS both the $n=4$ and $n=3$ poles are subtracted from each
bubble. The logarithmic divergence in $n=3$ corresponds to the
power-law divergence in $n=4$, and it then appears in the
expression for the bubble $I_0$. PDS thereby keeps track of linear
divergences (see Ref.~\cite{Ph99} for extensions of this idea).
This allows for ``fine-tuning'' between the coefficient $C_0$ and
the linear divergence.  In PDS, the integral $I_0$ in
Eq.~(\ref{eq:loopi}) is defined to be
\begin{eqnarray}
I_0^{\rm PDS}=-\left({M\over 4\pi}\right)\, (\mu + {\rm i}|{\bf k}| ),
\label{eq:ipdszero}
\end{eqnarray}
a very similar expression to that obtained using a momentum
cut-off, or with a momentum-space
subtraction~\cite{Ge98A,Ge98B,MS98}.

If PDS is employed the bubble sum (\ref{eq:geom}) yields
\begin{eqnarray}
T_{\rm NN} = {4\pi\over M}\  { 1 \over
{4\pi\over M C_0} + \mu + {\rm i} |{\bf k}|}.
\label{eq:bubsumLO}
\end{eqnarray}
Matching to the leading-order ERT amplitude (\ref{eq:TERE0}):
\begin{eqnarray}
{4\pi\over M C_0}\ =\
{1 \over a}-\mu.
\end{eqnarray}

PDS is designed to isolate the linear divergence in the integral
$I_0$. Ultimately this allows a portion of the attractive
potential represented by $C_0$ to cancel against the repulsive
effect of the virtual kinetic energy of the nucleons in the
potential well. The total effect of this kinetic energy is what is
calculated in $I_0$. If the size of the well is $R$ it scales as
$1/(RM)$: $R^3$ (for the total number of states in the well) and
$1/(MR^2)$ (Heisenberg uncertainty principle) for the kinetic
energy of particles in that well. By making $C_0$, which really
represents the depth of the well, also scale like $R/M$ we can
incorporate in our EFT the balance of potential and kinetic energy
that we know is necessary to produce a shallow bound state.  (See
Ref.~\cite{Sc97} for explicit implementations of this procedure
using co-ordinate space regulation of the potential
(\ref{eq:C0pot}).)  PDS incorporates exactly this cancellation,
but it also is more elegant than simply placing a co-ordinate
space regulator at a distance $R$, since the linear divergence is
manifest within the framework of dimensional regularization.

The explicit introduction of the subtraction scale $\mu$ into the
expression for the amplitude results in $C_0$ having its scale set by
$\mu$ ($\sim 1/R$) and not by $1/a$. Ultimately what this analysis
shows is that in order for the four-nucleon interaction in
Eq.~(\ref{eq:invar}) to generate a system with a shallow bound state
(i.e. $1/a \sim 0$), $C_0$ must be near a non-trivial fixed point. The
existence of this non-trivial fixed point~\cite{Ka98A,Ka98B} is
independent of the particular regularization and renormalization
scheme chosen~\cite{Bi99}, and its presence is what modifies the the
scaling of $C_0$ from the NDA ansatz (\ref{eq:NDA}). Turning the
argument of this section around we realize that {\it if} we wish to
construct a Lagrangian (\ref{eq:invar}) which reproduces the
low-energy bound state that is present in the NN system {\it then}
there must be a non-trivial fixed point in the flow of the coefficient
in the LO interaction Lagrangian.

Including higher-order terms in the EFT in the presence of such a
fixed point is discussed in detail in
Refs.~\cite{Ka98A,Ka98B,Ch99,Be00}. Here I will only say that to
move beyond the leading-order ERE amplitude we design our EFT to
reproduce the amplitude (\ref{eq:TERE2}) re-expanded as:
\begin{eqnarray}
T_{\rm NN}^{{\rm ERE} (2)}&=&-{4 \pi \over M}{1 \over -\frac{1}{a} + \frac{1}{2}
r_0 k^2 - {\rm i} k} \nonumber\\
&=&{4 \pi a \over M} \left[1 + {\rm i} a k - \frac{1}{2} a r_0 k^2
{1 \over 1 + {\rm i} a k} \right]^{-1} \nonumber\\
&=&{4 \pi a \over M}{1 \over 1 + {\rm i} a k} \left[1 + \frac{1}{2} a r_0 k^2
{1 \over 1 + {\rm i} a k} + \cdots \right].
\label{eq:EREreexpand}
\end{eqnarray}
In doing this we are assuming that $ak \sim 1$ but $a r_0 k^2 < |1
+ {\rm i} a k|$. In fact this second condition is satisfied as long as
$k < 1/r_0$. This technique works quite well in both the
${}^3{\rm S}_1$ and $\si$ NN channels. Thus ultimately the scale hierarchy
in the theory is:
\begin{equation}
k \sim 1/a \equiv Q \ll \Lambda \equiv 1/r_0 \sim m_\pi.
\end{equation}
$\nopi$ expands low-energy NN observables in powers of $Q/\Lambda$,
an expansion known as ``Q-counting''.

\subsection{The deuteron in $\nopi$}

\begin{figure}[t]
\centerline{{\epsfxsize=4.4in \epsfbox{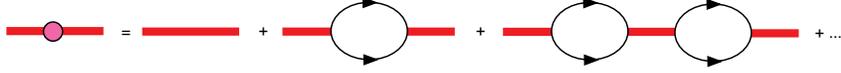}} }
\vskip -5 mm
\caption{The bare dibaryon propagator
is dressed by nucleon bubbles to all orders, resulting in a dressed
dibaryon propagator. Figure from Ref.~\cite{BS00}, courtesy S.~Beane
and M.~Savage.}
\label{fig-dressthetransvestite}
\vskip .2in
\end{figure}

We will now see that a re-organization of $\nopi$ is useful if we
wish to study deuteron properties.  To do this we introduce an
additional dibaryon field into theory, which we denote as $t$. The
$\nopi$ Lagrangian (\ref{eq:nopi1}) + (\ref{eq:invar}) can then be
rewritten as:
\begin{equation}
{\cal L}= N^\dagger \left[{\rm i} \partial_0 + \frac{\nabla^2}{2M}\right]N
+ t^\dagger \Delta t - y [t^\dagger NN + t N^\dagger N^\dagger].
\label{eq:Ltransvestite}
\end{equation}
Note that the dimensions of the $t$ field here are $[t]=3/2$.
This theory, or a close relative thereof, was proposed for NN
scattering in Ref.~\cite{Ka97}, but its antecedents are work due
to Weinberg~\cite{We63} and the Lee model. The degrees of freedom
are nucleons and the dibaryon field $t$. As in the Lee model there
is are vertices for $NN \leftrightarrow t$. Hence the dibaryon
field must be dressed by $NN$ loops (see
Fig.~\ref{fig-dressthetransvestite}). The Dyson equation for the
dibaryon propagator $D_t$ is then:
\begin{equation}
D_t={{\rm i} \over \Delta} + {{\rm i} \over \Delta} (-{\rm i} \Sigma_t) D_t,
\end{equation}
with $\Sigma_t$ being able to be evaluated exactly in this
theory:
\begin{equation}
\Sigma_t=y^2 I_0.
\end{equation}
If MS is used to regulate $I_0$ we discover that the $t$-field
propagator for a dibaryon of energy $E$ and momentum ${\bf 0}$ is
just:
\begin{equation}
D_t(E;{\bf 0})={{\rm i} \over \Delta + y^2 \frac{{\rm i} M k}{4 \pi}},
\label{eq:Dt}
\end{equation}
with $k^2=ME$. Employing the relation:
\begin{equation}
(-{\rm i}) T_{\rm NN}=y^2 D_t
\end{equation}
we find that the $T_{\rm NN}$ of Eq.~(\ref{eq:TERE0}) is recovered provided
that the identification
\begin{equation}
\left[{\Delta \over y^2}\right]^{\rm MS}={M \over4 \pi a}
\end{equation}
is made. Note that this also shows that the theory
(\ref{eq:Ltransvestite}) is equivalent to that (\ref{eq:invar}) of
the previous subsection, provided the appropriate identification
of $\Delta$ and $y^2$ with $C_0$ is made.  This can also be shown
by constructing the path-integral for (\ref{eq:Ltransvestite}) and
integrating over the field $t$ to obtain the Lagrangian
(\ref{eq:nopi1}) + (\ref{eq:invar})~\cite{BG99}.

\vskip 2 mm

\noindent (A technical note: it is, in fact, phenomenologically
advantageous to adjust:
\begin{equation}
\left[{\Delta \over y^2}\right]^{\rm MS}={M \gamma \over4 \pi},
\end{equation}
in the ${}^3{\rm S}_1$ channel. Here $\gamma=\sqrt{MB}$ is the
binding momentum of deuterium. Numerically $\gamma^{-1}=4.319$ fm.
By making this choice we ensure that the deuteron pole is at the
correct energy.)

%

A quick glance at the Lagrangian (\ref{eq:Ltransvestite}) reveals
that the field $t$ is not dynamic: it corresponds to a field of
infinite mass, since it has no energy terms. If we extend the
Lagrangian to include such a term:
\begin{equation}
{\cal L}=N^\dagger \left[{\rm i} \partial_0 + \frac{\nabla^2}{2M}\right]N
- t^\dagger \left[{\rm i} \partial_0 + \frac{\nabla^2}{4M} - \Delta \right]t
- y [t^\dagger NN + t N^\dagger N^\dagger],
\label{eq:Ltransvestite2}
\end{equation}
we observe that the field $t$ actually has the wrong sign for its
energy terms. This is a result of its auxiliary field nature.
Repeating the derivation of $T_{\rm NN}$ from the previous paragraph we
now obtain (using MS for the loop):
\begin{equation}
T_{\rm NN}(E;{\bf 0})={1 \over {\Delta \over y^2} - {k^2 \over M y^2}
+ {i M k \over 4 \pi}}.
\label{eq:Ttransmobile}
\end{equation}
This gives us the opportunity to adjust the two independent
parameters $y$ and $\Delta$ to reproduce not just the first, but
also the second, term in the effective range expansion for
$T_{\rm NN}$, i.~e. to reproduce Eq.~(\ref{eq:TERE2}).

Note that if we want to have both $1/a$ and $\frac{1}{2} r_0 k^2$
in the denominator of our expression for $T_{\rm NN}$ then it must
be that both are of the same order in our power counting.
Otherwise we could re-expand the second-order effective-range
amplitude as per Eq.~(\ref{eq:EREreexpand}). If we wish to have
$1/a \sim \frac{1}{2} r_0 k^2$, and $1/a$ and $k$ are both
low-energy scales $\sim Q$ then the logical conclusion is that
$1/r_0$ is also a low-energy scale $\sim Q$.  In other words, both
$a$ and $r_0$ are unnaturally large and this should be taken into
account in building the EFT~\cite{BS00}.

Simple algebra then reveals that the choices:
\begin{equation}
y^2={8 \pi \over M^2 r_0}; \qquad \Delta^{\rm MS}={2 \over M r_0 a},
\end{equation}
will turn Eq.~(\ref{eq:Ttransmobile}) into Eq.~(\ref{eq:TERE2}).  Note
that if $1/r_0$ scales as $Q$, then $y \sim Q^{1/2}$.

\begin{exercise}
Show that if PDS is used to regulate $I_0$ the value
of $y^2$ does not change, but:
\begin{equation}
\Delta^{\rm PDS}={2 \over M r_0}\left({1 \over a} - \mu \right).
\end{equation}
\end{exercise}

In retrospect it is better to write the Lagrangian for this theory in
terms of a rescaled field $\tilde{t}=y t$. Then:
\begin{equation}
{\cal L}=N^\dagger \left[{\rm i} \partial_0 + \frac{\nabla^2}{2M}\right]N -
\frac{1}{y^2} \tilde{t}^\dagger \left[{\rm i} \partial_0 +
\frac{\nabla^2}{4M} - \Delta \right]\tilde{t} - [\tilde{t}^\dagger NN +
\tilde{t} N^\dagger N^\dagger].
\end{equation}
The dibaryon field $\tilde{t}$ now scales with the small parameter $Q$ as
$\tilde{t} \sim Q^{1/2} Q^{3/2}=Q^2$.

In the ${}^3{\rm S}_1$ channel the dibaryon is a field for the
deuteron, and the deuteron is ``dressed'' by $NN$ loops, which
simply means that unitarity is incorporated in the theory. In the
case of the deuteron it makes more sense to rewrite
Eq.~(\ref{eq:TERE2}) as
\begin{equation}
T={4 \pi \over M}{Z_{\rm d} \over \gamma + {\rm i} k} + {\cal R}(k),
\label{eq:Td}
\end{equation}
where, in the language of quantum field theory $Z_{\rm d}$ is
related to the wave-function renormalization of the deuteron field
$\tilde{t}$. Meanwhile $k=i \gamma$ is the deuteron pole position
in the complex $k$-plane, and represents the ``typical'' momentum
scale inside deuterium.

\begin{exercise}
Show that the second-order ERE amplitude (\ref{eq:TERE2}) leads to
\begin{equation}
Z_{\rm d}=\(1 - {2 r_0 \over a}\)^{-1/2}=1.64.
\label{eq:Zd}
\end{equation}
\end{exercise}

\subsubsection*{Hand-waving about power counting}

Effects due to higher-order terms in the ERE, mixing with the
$\diii$ state, etc.  can be systematically incorporated in the EFT
(\ref{eq:Ltransvestite2}). Here I focus on power counting for the
coupling of the deuteron to photons.

\begin{figure}[t]
\centerline{{\epsfxsize=2.8in \epsfbox{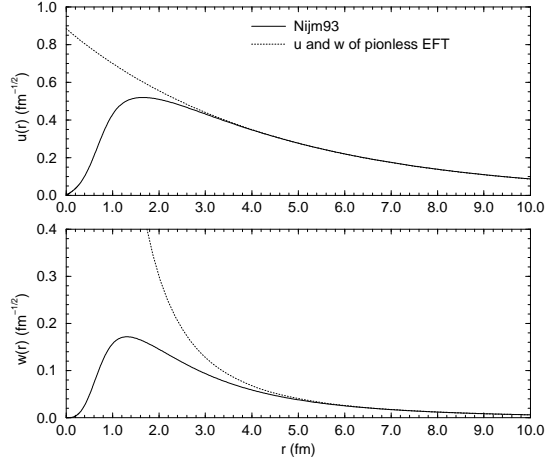}} }
\vspace{-0.7cm}
\caption{Radial S-wave and D-wave wave deuteron
wave functions for $\nopi$ (or for ERT). Also shown is the wave
function obtained from the Nijm 93 potential model~\cite{St94}.}
\label{fig-nopiwf}
\end{figure}

The single-nucleon coupling to photons is mandated by gauge
symmetry and $\gamma$N data. Meanwhile the unknown two-nucleon
physics manifests itself in the theory as short-distance couplings
of the photon to the dibaryon, for instance $t^\dagger \nabla^2
A_0 t$.  Such effects are suppressed because ``most'' of the
deuteron wave function ``lives'' at long-distances (see
Fig.~\ref{fig-nopiwf}). Beginning with Eq.~(\ref{eq:Td}) and
taking appropriate Fourier transforms one can derive the $\nopi$
deuteron radial S-wave wave function as:
\begin{equation}
u(r)=\sqrt{2 \gamma Z_{\rm d}}\, \exp(- \gamma r) \, .
\label{eq:uas}
\end{equation}
This is, of course, exactly the same as the effective range theory
(ERT) wave function for this state. It stems from the idea that
the long-range part of the deuteron wave function is prescribed by
two numbers, the deuteron binding energy, which determines the
exponential fall-off at long distances, and the coefficient of the
exponential, which is known as $A_{\rm S}$ in the literature and
has an ``experimental'' value~\cite{dS95}
\begin{equation}
A_{\rm S}=0.8845 \pm 0.0008~{\rm fm}^{-1/2}.
\end{equation}
This compares favourably with the ERT/$\nopi$ prediction from
Eq.~(\ref{eq:Zd}): $A_{\rm S}=0.871~{\rm fm}^{-1/2}$. Effective
range theory is an accurate way to extrapolate the NN amplitude
from the physical region to the deuteron pole.

In ERT, or in $\nopi$, this ``asymptotic'' wave function persists
all the way in to $r=0$, since in $\nopi$ the potentials involved
are all zero range, and so nothing modifies the asymptotic form
(\ref{eq:uas}) until the origin is reached.  We would expect that
the asymptotic form would start to deviate from more detailed wave
functions at $r \sim 1/\Lambda= 1/m_\pi$, since that is the range
of the underlying potential involved in NN physics. Thus the
amount of wave function at short distances is parametrically
small, being only $\gamma/\Lambda$ of the total. However, in ERT
there is no way to correct for this error, and so predictions of
ERT should in principle be valid only up to corrections of
relative size $\gamma/\Lambda \sim 1/3$.

In the next subsection we will use the calculation of the charge
form factor of deuterium to show that in $\nopi$ we can introduce
operators which systematically correct for the (brutal)
approximation that the form (\ref{eq:uas}) persists all the way to
$r=0$. These operators are generically suppressed by (at least)
one factor of the expansion parameter
\begin{equation}
Q \equiv {(q,\gamma) \over \Lambda},
\end{equation}
where $q$ is the momentum of the external (electromagnetic or
weak) probe of the deuteron. $Q$ is thus the expansion parameter
for calculations of processes on deuterium in $\nopi$. For any
given reaction the effects that can contribute are split up into
``long-distance'' parts due to single nucleons interacting with an
external probe at distances $r \sim 1/\gamma, 1/q$, and
``short-distance'' parts due to the probe interacting with the
nucleons in the unknown region $r \sim 1/\Lambda$. The
``long-distance'' mechanisms can be calculated from knowledge of
single-nucleon interactions with the external probes, and because
of the ``fluffy'' nature of the deuteron wave function (see
Fig.~\ref{fig-nopiwf}) the short-distance operators are always
suppressed relative to the long-distance ones. This gives $\nopi$
considerable predictive and explanatory power.

\subsection{Deuteron charge form factor}

The relevant Feynman rules for the calculation of the deuteron's
charge form factor are as follows:

\begin{center}
{\bf Feynman rules and power counting:}
\end{center}
\begin{enumerate}
\item With each two-nucleon propagator we associate a factor:
\begin{eqnarray*}
{{\rm i} \over E - {\bf p}^2/M},
\end{eqnarray*}
where $E$ is the energy of the NN pair and ${\bf p}$ is the
relative momentum. (This two-nucleon Green's function results from
doing the relative-energy integration for the product of two
one-nucleon propagators.)  Since $E=-B=-\gamma^2/M$ both terms in
the denominator scale with $Q$ as $Q^2/M$, and so this propagator
is assigned a factor $Q^{-2}$.

\item With each dibaryon-nucleon-nucleon vertex we associate a
factor $(-{\rm i})$. Clearly this carries no powers of the small
momentum $Q$, and so  scales as $Q^0$.

\item Each loop corresponds to an integral over the loop
three-momentum. (The zeroth-component integration has already been
performed, see rule 1.) This results in a factor of $Q^3$.

\item The interaction of an $A_0$ photon with a single nucleon
that results from the use of the minimal substitution $\partial_0
\rightarrow \partial_0 - {\rm i} Q A_0$ yields a $\gamma NN$ vertex
corresponding to a factor $-{\rm i} e$. We will count factors of $e$
separately here, and this vertex then scales as $e$ (surprise).

\item The same minimal substitution in (\ref{eq:Ltransvestite2})
yields a $\gamma \tilde{t} \tilde{t}$ vertex ${\rm i} e/y^2$.
Since $y \sim Q^{1/2}$ this coupling scales as $e/Q$ and so is
``super-leading'', being enhanced over the coupling of a photon to
a single nucleon. As will see this super-leading behaviour is a
necessary consequence of electromagnetic current conservation.

\item Other electromagnetic vertices not produced by the minimal
substitution procedure are also permitted in the theory. One such
vertex involves a higher-derivative interaction of an $A_0$ photon
with a single-nucleon. This may be represented as:
\begin{equation}
{\cal L}_{\mbox{finite nucleon size}}=-{e \langle r_{\rm N}^2
\rangle \over 6}\, N^\dagger {\nabla^2} A_0 N \, .
\label{eq:finsize}
\end{equation}
This generates a higher-dimension $\gamma NN$ vertex $-{\rm i} e
\langle r_{\rm N}^2 \rangle {\bf q}^2/6$. This scales as $e Q^2$.
(In fact, it is even more suppressed than this might indicate,
since the isoscalar radius of the nucleon is not set by the pion
mass, but instead by dynamics at the scale of chiral symmetry
breaking $\Lambda_\chi$.)

\item Analogously, there are vertices of the form $e \xi
\tilde{t}^\dagger \nabla^2 A_0 \tilde{t}$. Naive dimensional
analysis, together with the knowledge that $[\tilde{t}]=2$,
suggests that $[\xi]=-3$. Hence we employ NDA to write $\xi \sim
1/\Lambda^3$, and so have this vertex scaling as $e Q^3$. The
extra power of $Q$ as compared to the scaling of the operator in
(\ref{eq:finsize}) can be thought of as coming from the diffuse
nature (technically known as ``fluffiness'') of the deuteron wave
function.
\end{enumerate}

We are now in a position to calculate the charge form factor of
the deuteron, $G_{\rm C}$. In terms of standard quantum mechanics the
charge form factor of the deuteron is defined by:
\begin{eqnarray}
G_{\rm C}&=&{1 \over 3 e} \left(
\left \langle 1\left|J^0\right|1 \right \rangle +
\left \langle 0\left|J^0\right|0 \right \rangle +
\left \langle -1\left|J^0\right|-1 \right \rangle
\right),
\label{eq:GC}
\end{eqnarray}
where we have labeled the (non-relativistic) deuteron states by
the projection of the deuteron spin along the direction of the
momentum transfer ${\bf q}$ and $\eta \equiv |{\bf q}|^2/(4 M_{\rm
d}^2)$.  Here it suffices to say that $G_{\rm C}(Q^2)$ represents
the amplitude for coupling of a ``Coulomb'' photon with
four-momentum $(0,{\bf q})$, $Q^2={\bf q}^2$ to couple to the
deuteron. Hence it is calculated by computing the amplitude for
coupling of an $A_0$ photon to the deuteron.

\begin{figure}[!ht]
\centerline{{\epsfxsize=3.3in \epsfbox{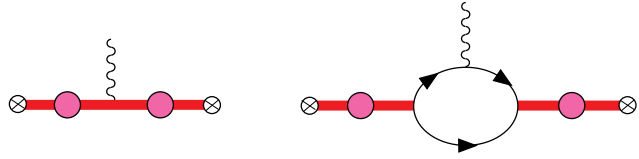}} }
\vskip -5 mm
\caption{The leading-order diagrams that contribute to the deuteron
charge form factor. The thick solid lines represent the deuteron field,
while the thin lines with arrows denote the nucleon field. The crosses
indicate the need for wave-function renormalization. Figure from
Ref.~\cite{BS00}, courtesy S.~Beane and M.~Savage.}
\label{fig-GCdiags}
\end{figure}

The ``leading-order'' graphs for this process are depicted in
Fig.~\ref{fig-GCdiags}. The graph on the right contains two
two-nucleon propagators, one loop integral, and one $\gamma$NN
vertex $-{\rm i} e$. Hence it scales as:
\begin{eqnarray*}
{Q^3 \over Q^2 Q^2}\, e \sim {e \over Q}.
\end{eqnarray*}
Meanwhile the diagram on the left includes only the vertex $i
e/y^2$ and so also scales as $e/Q$. Together they produce a
deuteron form factor:
\begin{equation}
-{\rm i} e G_{\rm C}=y^2 \gamma r_0 Z_{\rm d}
\left[(-{\rm i})^2 \int {{\rm d}^3 p
\over (2 \pi)^3} {{\rm i} \over -B - {\bf p}^2/M}
(-{\rm i}e) {{\rm i} \over -B - \left({\bf p}
+ {\bf q}/2\right)^2/M} + {{\rm i} e \over y^2} \right] \, ,
\label{eq:GCexp}
\end{equation}
where we have neglected deuteron recoil, and incorporated an overall
factor of $y^2 \gamma r_0 Z_{\rm d}$ in order to account for the
wave-function renormalization that turns the amplitude for the $A_0
\tilde{t} \tilde{t}$ interaction into the amplitude for an $A_0$ photon
coupling to the deuteron. (For details on wave-function
renormalization in this approach see Refs.~\cite{Be00,BS00}.)

\begin{exercise}
Use the Fourier representation of $1/(MB + {\bf p}^2)$ in order to
convert the integral in Eq.~(\ref{eq:GCexp}) into an integral over
${\bf r}$. You should find that
\begin{equation}
G_{\rm C}=\gamma r_0 Z_{\rm d} \left[ {y^2 M^2 \over 16 \pi^2}
\int {\rm d}^3 r \, {\exp(-\gamma r) \over r}\,
 \exp({\rm i} {\bf q} \cdot {\bf r}/2)\,
{\exp(-\gamma r) \over r}  - 1 \right] \, .
\end{equation}
Note that the first term here can be interpreted as
\begin{equation}
\langle \psi|\exp({\rm i} {\bf q} \cdot {\bf r}/2)|\psi \rangle,
\end{equation}
where $|\psi  \rangle$ is the deuteron wave function corresponding
to the radial wave function (\ref{eq:uas}). Evaluate the
co-ordinate space integral to obtain:
\begin{equation}
G_{\rm C}={\gamma r_0 \over 1 - \gamma r_0}
\left[{4 \over q r_0} \arctan\left({q \over 4 \gamma}\right) - 1\right].
\label{eq:GCNLO}
\end{equation}
\end{exercise}

The first correction to this result is due to the single-nucleon
operator from Eq.~(\ref{eq:finsize}). It gives an effect of
$O(Q^2)$ relative to those included in (\ref{eq:GCNLO}). This
effect is easily included since the square of the nucleon's
isoscalar charge radius, $\langle r_{\rm N}^2 \rangle$, is known.
The result (\ref{eq:GCNLO}) is a prediction for $G_{\rm C}$ based
only on low-energy NN data ($r_0$ and $B$) and current
conservation. Current conservation imposes the constraint:
\begin{equation}
G_{\rm C}(q^2=0)=1.
\label{eq:betterbe}
\end{equation}
Note that without the leading-order effect of direct coupling of
the $A_0$ photon to the deuteron field this condition would have
been violated, since the right-hand diagram of
Fig.~\ref{fig-GCdiags} yields $Z_{\rm d}$ for the deuteron charge,
which is larger than one. This is just a consequence of the volume
integral of the probability density corresponding to
Eq.~(\ref{eq:uas}) being greater than one, but carefully
maintaining current conservation in $\nopi$ ensures that
Eq.~(\ref{eq:betterbe}) is satisfied.

Meanwhile the slope of $G_{\rm C}$ at $q^2=0$ gives $-\frac{1}{6}
\langle r_{\rm d}^2 \rangle$. In $\nopi$ we obtain the number:
\begin{equation}
\langle r_{\rm d}^2 \rangle={Z_{\rm d} \over 8 \gamma^2}=
 (1.990~\mbox{fm})^2.
\label{eq:nopird}
\end{equation}
There are corrections to $G_{\rm C}$, and hence to $\langle r_{\rm
d}^2 \rangle$, from short-distance operators involving $A_0$
couplings to $t^\dagger t$, i.~e.  interactions of the last type
discussed in the Feynman rules listed above. These operators scale
like $q^2$, while leading-order prediction scales like $1/Q$, thus
we expect that the short-distance effects that are the first
correction to (\ref{eq:nopird}) to be down by a relative factor of
$\gamma q^2 \sim Q^3$. This means that the prediction
(\ref{eq:nopird}) should be accurate to 3\%. And, indeed this is
the case with the difference between (\ref{eq:nopird}) and the
experimental number $\langle r_{\rm d}^2
\rangle=(1.971~\mbox{fm})^2$. being on the order of 2\%.

Although the result (\ref{eq:nopird}) can be derived in ERT too,
systematic error estimates such as these are not possible in ERT.
Also, it is not entirely clear how to include finite-nucleon-size
effects in ERT, and the ERT wave function (\ref{eq:uas}) will
violate the current-conservation condition (\ref{eq:betterbe}). As
presented here, $\nopi$ is a descendant of Bethe's work, but it
has evolved so that it has features that make it more suitable for
the accurate computation of low-energy electromagnetic observables
in the two-nucleon system.

\begin{figure}[t]
\centerline{{\epsfxsize=3.5in \epsfbox{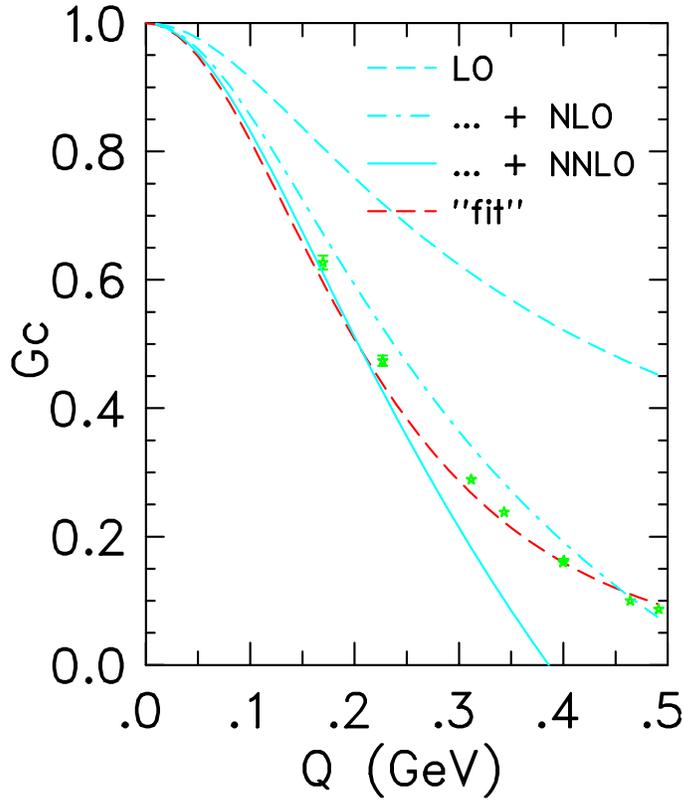}}}
\vskip -5 mm
\caption{A comparison of $G_{\rm C}$ from $\nopi$~\cite{Ph99,BS00} with a 
``fit'' function (long-dashed line) and experimental data points. The
$\nopi$ calculation described above is represented by the dot-dashed
line (i.~e. it is NLO in the terminology of this figure), while the
inclusion of finite-size effects for the nucleon then produces the,
solid, ``NNLO'' curve, which is accurate up to corrections at relative
order $O(e Q^2)$ such as those due to short-distance effects in the
deuteron wave function. Figure courtesy R.~Gilman, from
Ref.~\cite{GG01}.}
\label{fig-GClowq}
\end{figure}

For what it is worth I have included a comparison of the $\nopi$
prediction discussed here, and computed in Refs.~\cite{Ph99,BS00},
with the (sparse) low-$q^2$ data on $G_{\rm C}$
(Fig.~\ref{fig-GClowq}). We see that the $\nopi$ calculation breaks
down at momenta $q \sim 2 m_\pi$, which is exactly what we would
expect from the fact that at such momenta we begin to probe relative
separations of the neutron and the proton where the theory completely
fails to describe the dynamics~\footnote{The factor of two in 
``$q \sim 2 m_\pi$'' arises because only half of the virtual photon's
momentum is transferred to the deuteron's relative degree of
freedom. The other half is deposited into motion of the deuteron's
centre of mass.}.  In this regime we are no longer in the
long-wavelength limit where the pion dynamics can be ignored. In
Section~\ref{sec-chipt} we will discuss how to improve the description
of the deuteron state $|\psi \rangle$ in this important regime.

While I do not have space here to do justice to the literature on
$\nopi$ it is worth mentioning that a much more impressive and
important application of $\nopi$ to low-energy electromagnetic
processes involving the deuteron is the 1\% calculation of the
process ${\rm np} \rightarrow {\rm d} \gamma$. $\nopi$ allows the
derivation of analytic expressions for the E1 and M1 isovector
strengths that dominate this process at low photon energies. Using
the formalism discussed here, or the equivalent approach adopted
by Rupak in his important calculation~\cite{Ru99} we can derive
results for $\gamma {\rm d} \leftrightarrow {\rm np}$ that agree
with the experimental data on low-energy deuteron
photodisintegration at the expected level of accuracy. This is
useful not only for what it teaches us about $\gamma {\rm d}
\rightarrow {\rm np}$, but also because it facilitates a
computation of radiative neutron capture at the photon energies of
relevance to big-bang nucleosynthesis by using a framework in
which errors can be reliably estimated and analytic expressions
developed. Similar work on $\nu d$ interactions in $\nopi$ by
Butler, Chen, and Kong, resulted in analytic expressions for the
process $\nu_e {\rm d} \rightarrow {\rm e}^- {\rm pp}$ that were
then used in computing the deuteron breakup rates critical to the
assessment of the SNO data on neutrino oscillations~\cite{Bu01}
and also in calculating the solar hep process ${\rm pp}
\rightarrow {\rm d e}^+ \nu_e$~\cite{BC01}.  Finally, as a new
generation of Compton scattering experiments on deuterium come on
line $\nopi$ holds out the promise of empowering model-independent
extractions of the neutron polarizability from the reaction
$\gamma {\rm d} \rightarrow \gamma {\rm d}$~\cite{GR01}. (See also
below.)

To conclude this section, let me say that $\nopi$ as applied to
processes on deuterium computes the ``long-distance'' portion of
observables using wave functions that could have been written down
using effective-range theory over 50 years ago. However, the
errors in these wave functions can be systematically corrected for
provided the probe is not of sufficient energy to truly ``see''
the ``short-distance'' part $r \leq 1/m_\pi$ of the NN
interaction. Thus, $\nopi$ is clearly a descendant of ERT, but it
improves on it, since it allows for systematically-improvable
calculations of (very) low-energy processes on the deuteron, and
these resultant calculations have a well-justified theoretical
error bar. This systematicity arises the existence of the
small(ish) parameter
\begin{equation}
Q={\gamma \over m_\pi}.
\end{equation}
Such mingling of old and new to the benefit of nuclear physics is,
I think, quite nice. To quote a very old source~\cite{Matt}, he (or she)
who applies these ideas

\begin{quotation}
\ldots is like the owner of a house who brings out of his
storeroom new treasures as well as old.
\end{quotation}

\section{The three-body system with short-range interactions}

In this section we turn our attention to the three-body problem in
$\nopi$. The problem to be solved is that of three particles
interacting via forces that are short-ranged compared to the
distance scales of interest. If these particles have spin-zero
then one system where this physical condition is satisfied is the
so-called helium trimer. Here the scattering length for two helium
atoms is,
\begin{equation}
a_2 \approx 125~\AA,
\label{eq:a2}
\end{equation}
and these two helium atoms can come together to call a He-He bound
state known as a dimer. Complicated Faddeev calculations suggest
that the (three-body) scattering length for dimer-atom scattering
is:
\begin{equation}
a_3 \approx 195~\AA.
\label{eq:a3}
\end{equation}
Both $a_2$ and $a_3$ are much larger than the typical range of the
He-He interaction, which is about $10~\AA$ or so.

As nuclear physicists we are also interested in the problem of
neutron-deuteron scattering~\footnote{There is, of course, more
proton-deuteron data, but in that case Coulombic effects somewhat
mask the strong interaction effects we are trying to work on here.
For a first effort at including Coulomb corrections in EFTs of the
3N system, see Ref.~\cite{RK01}.}. Here the relevant scales
are~\cite{dS95,Gi92}:
\begin{equation}
a_2=5.4194 \pm 0.0020~{\rm fm}; \, \, \,a_3^{(3/2)}=6.45 \pm 0.02~{\rm
fm}; \, \, \, a_3^{(1/2)}=0.65 \pm 0.04~{\rm fm},
\end{equation}
where $a_2$ is the ${}^3{\rm S}_1$ scattering length. There are
two three-body scattering lengths because the neutron is a
spin-half particle, and the deuteron a spin-one particle, and so
there are two possible channels for nd scattering: the $S=3/2$
(quartet) and $S=1/2$ (doublet), with $S$ the total spin of the nd
system.  Since the range of the NN interaction is still $R \sim
1/m_\pi \sim 1.5$ fm, we hope and believe that $\nopi$ will be a
viable way to treat 3N systems in low momentum reactions.

As I will explain here, attacking the quartet channel in $\nopi$
turns out to yield remarkably accurate predictions for low-energy
nd scattering without very much work. However, the doublet channel
is much trickier, and significant conceptual work on the nature of
renormalization in this problem was required before $\nopi$ could
be made to yield answers. In fact, the conceptual problem in the
$S=1/2$ channel also arises when discussing the scattering of
three spinless bosons (such as helium atoms), and so, in order to
avoid complications with spin, the helium-atom problem is the one
we will discuss here. The good news is that this conceptual work
has been (largely) done and there is now the promise of
calculations akin to those described in the previous section, only
for 3N systems. The discussion in this section mirrors closely
that in the original papers on $\nopi$ applied to the three-body
system. See especially Refs.~\cite{BvK97,Bd98,Bd99A,Bd99B}.

\subsection{The equation}

We wish to describe the scattering of three spinless bosons via
short-range interactions. Two of the bosons can form a bound
state, which we include in the theory via an additional field $t$
(``dimer'').  The Lagrangian for the theory is then exactly that
of Eq.~(\ref{eq:Ltransvestite}) only with an additional term
representing dimer-boson interactions:
\begin{equation}
{\cal L}=N^\dagger \left[{\rm i} \partial_0 + \frac{\nabla^2}{2M}\right]N +
t^\dagger \Delta t - y [t^\dagger NN + t
N^\dagger N^\dagger] - h \psi^\dagger \psi t^\dagger t.
\label{eq:3Btransvestite}
\end{equation}
(Note that due to a peculiar notational choice ``$N$'' now
represents a Helium atom. Note also that $\tilde{t}$ has been
discarded in favour of the original $t$.)  We assume that all
other higher-dimensional operators are indeed suppressed, and so
work within the theory defined by Eq.~(\ref{eq:3Btransvestite}).
Furthermore, remembering that $t$ is a dimension $3/2$ field, we
realize that $[h]=-2$, so naive dimensional analysis implies that
the three-body force term should be suppressed, and we thus expect
to be able to set $h=0$ to get the leading-order result. We shall
initially proceed in this way, but it is worth warning that we
have already seen that in the presence of ``unnaturally'' long
distance scales, such as the scattering length (\ref{eq:a2}), NDA
is too naive.

\begin{figure}[t]
\centerline{\psfig{file=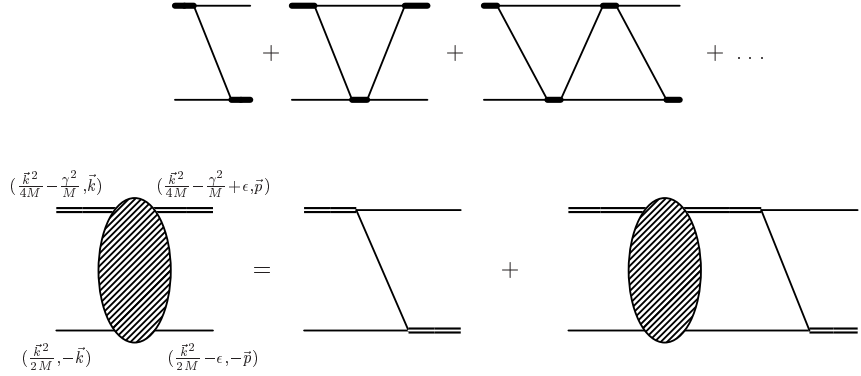,height=2.3in,
             bbllx=70pt,bblly=560pt,bburx=520pt,bbury=770pt}}
\vskip -5 mm
\caption{The first line shows the infinite sum of
graphs giving the leading order neutron-deuteron scattering in the
spin $3/2$ channels. The second line is the graphical
representation of the Faddeev equation that sums them up. Figure
from Ref.~\cite{Be00}.}
\label{fig-3Bheq0}
\end{figure}

Now consider boson-dimer scattering within this theory. The incoming
dimer is the dressed field, with the propagator given by
Eq.~(\ref{eq:Dt}).  All of the graphs that contribute to boson-dimer
scattering in the theory with $h=0$ are shown on the top line of
Fig.~\ref{fig-3Bheq0}.  These graphs just form the multiple-scattering
series for the three-body problem with short-range interactions, and
can be re-summed by writing the boson-dimer scattering amplitude as
the solution of a Faddeev equation. Indeed, in this case, since the
potential is separable the Faddeev equation reduces to a
Lippmann-Schwinger equation with one-nucleon exchange as the
potential (bottom line of Fig.~\ref{fig-3Bheq0}).

This can be understood from an EFT point of view by simple power
counting.  Each additional loop in one of the graphs on the top
line of Fig.~\ref{fig-3Bheq0} adds two $tNN$ vertices (each $\sim
y$), one extra non-relativistic three-nucleon propagator ($\sim
M/Q^2$), and one extra $tN$ propagator ($\sim 1/(M y^2 Q)$).
Including the three-dimensional loop integration we see that the
overall effect is that a loop is down by:
\begin{equation}
y^2 \times {M \over Q^2} \times {1 \over M y^2 Q} \times Q^3 \sim Q^0,
\end{equation}
i.~e. it is not suppressed at all. Thus it is necessary to sum
this entire class of loop graphs just to get the LO answer for the
boson-dimer scattering amplitude.

To derive this equation we choose the kinematics shown in
Fig.~\ref{fig-3Bheq0}.  The outgoing (incoming) dimer field
carries momentum ${\bf k}$ (${\bf p}$). Meanwhile, the outgoing
(incoming) boson carries momentum $-{\bf k}$ ($-{\bf p}$), The
outgoing deuteron and nucleon are on shell, and $\epsilon$ denotes
how far the incoming particles are off shell. This is the
``half-off-shell'' amplitude for $tN$ scattering.  We now let
$t({\bf p},0)$ be the amplitude for the sum of all Feynman graphs
in the theory (\ref{eq:Ltransvestite}) which contribute to $tN$
scattering in this kinematics (with $h=0$ here):

The integral equation for $t$ can then be read off from
Fig.~\ref{fig-3Bheq0}. At ${\bf k}=0$ it is:
\begin{eqnarray}
  \label{faddeevequation}
t({\bf p},0)={-M \over {\bf p}^2 + \gamma^2}
+ 8 \pi \lambda \int {{\rm d}^3 q \over (2 \pi)^3}\,
{1 \over \gamma^2 + {\bf p}^2 + {\bf q}^2 + {\bf p} \cdot {\bf q}}\,
{t({\bf q},0) \over \sqrt{\frac{3}{4}{\bf q}^2 + \gamma^2} - \gamma}
 \, ,
\label{eq:vecequation}
\end{eqnarray}
where the artificial parameter $\lambda$ has been introduced, with
$\lambda=1$ for the case of boson-dimer scattering.

\begin{exercise}
Check that applying the Feynman rules for the field theory
(\ref{eq:Ltransvestite}) leads to Eq.~(\ref{eq:vecequation}).
\end{exercise}

At zero relative momentum only S-waves contribute and so we
postulate: an amplitude
\begin{equation}
t({\bf p},0)=-{3M \over 8}\,
{1 \over \gamma + \sqrt{\gamma^2 + \frac{3}{4} p^2}}\, a(p)
\equiv -{M \over F(p)}\, a(p).
\end{equation}
Working with $a(p)$ rather than $t(p,0)$ makes our life easier
when we come to compare with experiment, since $a(0)=-a_3$, the
negative of the three-body scattering length.

Equation (\ref{eq:vecequation}) now boils down to a one-dimensional
integral equation for $a(p)$:
\begin{equation}
{1 \over F(p)}\, a(p)={1 \over p^2 + \gamma^2}
+ {2 \lambda \over \pi} \log
\int {\rm d}q\, {1 \over 2pq} \log\left({p^2 + q^2 + \gamma^2 +
pq \over p^2 + q^2 + \gamma^2 - pq}\right)\, a(q) \, .
\label{eq:aeq}
\end{equation}
This is the Faddeev equation for S-wave scattering for the case of
contact forces derived firstly in reference ~\cite{STM57} by
different methods. (Note: in fact, it was actually derived before
Faddeev derived his equations!)  Eq.~(\ref{eq:aeq}) can now easily
be solved numerically.

\begin{exercise}
Derive Eq.~(\ref{eq:aeq}) from Eq.~(\ref{eq:vecequation}).
\end{exercise}

\subsection{The problem}

Our next goal is to derive the ultraviolet (i.e. high momentum)
behaviour of the half-off-shell amplitude $a(p)$. For $p \gg
\gamma$ the second term on the right-hand side of
Eq.~(\ref{eq:aeq}) dominates, and the integral equation becomes:
\begin{equation}
a(p)={4 \lambda \over \sqrt{3} \pi}\,
\int_0^\Lambda {{\rm d}q \over q}\,  \log
\left({p^2+p q + q^2 \over p^2-p q + q^2} \right).
\label{eq:UV}
\end{equation}
A few things about this equation are worth noting:
\begin{enumerate}
\item A cutoff has been applied, in order to regulate otherwise
troublesome behaviour from the high-momentum part of the integral.

\item This is a {\it homogeneous} integral equation. No
information from the driving (inhomogeneous) term survives in the
UV limit.  Some authors have argued that this makes the original
problem defined by the equation (\ref{eq:aeq}) ill-posed, since
the kernel is non-compact~\cite{Ge00}, but in fact the kernel is
compact provided that any finite cutoff is imposed.

\item If $a(p)$ is a solution of Eq.~(\ref{eq:UV}) then
so is its complex conjugate $a^*(p)$.

\item {\it If} we were to take $\Lambda \rightarrow \infty$ then
the integral on the right-hand side is scale invariant.

\item Similarly, if $\Lambda \rightarrow \infty$ and $a(p)$ is a
solution then $a(1/p)$ is also a solution.
\end{enumerate}

This last point suggests that we try power-law solutions of the
form $a(p)=p^s$. Doing this, and performing the resulting Mellin
transform on the right-hand side, we reproduce an old result of
Danilov~\cite{Da63}: The homogeneous equation (\ref{eq:UV}) will
always have such a solution provided that $s$ obeys the
transcendental equation:
\begin{equation}
1- {8 \lambda \over \sqrt{3} s}\,
  {\sin(s\pi/6) \over \cos(s \pi/2)}=0 \, .
\label{eq:paulo}
\end{equation}
Now, the solution $a(p)$ has very different properties depending
on whether Eq.~(\ref{eq:paulo}) admits real or complex solutions
for $s$. Thus, we divide our discussion into two cases: $\lambda <
\lambda_{\rm c}=\frac{3 \sqrt{3}}{4 \pi}$, in which case the solution of
(\ref{eq:paulo}) is real, and $\lambda > \lambda_{\rm c}$, in which case
it is complex.  Note that the case we were originally considering
was $\lambda=1 > \lambda_{\rm c}$.

\subsubsection{The simple case: $\lambda < \lambda_{\rm c}$}

However, the case $\lambda < \lambda_{\rm c}$ is also interesting,
not least because it transpires that the equation (\ref{eq:aeq})
with $\lambda=-1/2$ describes neutron-deuteron scattering in the
quartet (spin-3/2) channel. In this case solutions to
Eq.~(\ref{eq:paulo}) are
\begin{equation}
s=\pm 2,\pm 2.17,\ldots.
\end{equation}
Here we discard the solutions that grow with $p$ as we go far
off-shell, and so obtain:
\begin{equation}
a(p)=C p^{-s}.
\end{equation}
It follows that the integral equation is well-behaved and the
results are completely insensitive to the size of the cutoff
$\Lambda$, and to the form of the cutoff function employed. (As
long, of course, as $\Lambda \gg \gamma$.) We could express this
schematically by saying that:
\begin{equation}
\Lambda \left. {{\rm d} a(p) \over {\rm d} \Lambda}
\right|_{\mbox{low p}} \approx 0 \, .
\end{equation}
This allows us to make a confident prediction for $a(p)$, and
hence for $a_3=-a(0)$. That prediction is that the quartet
neutron-deuteron scattering length is~\cite{STM57}:
\begin{equation}
a_3^{3/2}=5.09~{\rm fm}.
\label{eq:LO}
\end{equation}
However, we should remember that the deuteron propagator employed
to obtain this number reproduced only the deuteron binding energy.
Putting in a deuteron kinetic energy, allows us to also reproduce
$A_{\rm S}$, or, equivalently, $r_0$. If this is done, and the
field theory (\ref{eq:Ltransvestite2}) used to perform the above
analysis we instead find~\cite{BvK97}:
\begin{equation}
a_3^{3/2}=6.33 \pm 0.1~{\rm fm};
\label{eq:NLO}
\end{equation}
(See Ref.~\cite{Ef91} for an earlier, similar approach to this
problem.) Note that the correction due to these ``range effects''
is about 20\% of the leading-order $a_3^{3/2}$ (\ref{eq:LO}): a
number that might have been anticipated given the fact that
$\gamma r_0 \sim 0.4$. The error $\pm 0.1$ fm comes from using
analogous scale arguments to estimate the effect of including more
information about NN scattering in the deuteron propagator, as
well as what would happen if NNN forces were to be included in
this channel.

The number (\ref{eq:NLO}) compares well with the experimental
value and with sophisticated Faddeev calculations using modern,
accurate, NN potentials~\cite{Fr99}.  In fact, the calculation
(\ref{eq:NLO}) is model independent, and involves only the input
parameters $\gamma$ and $r_0$, thus one would expect {\it any}
potential model which reproduces these numbers to agree with the
result (\ref{eq:NLO}), within the estimated error.  It is also
important to note that model-independent statements of similar
accuracy can be made for the energy-dependence of the quartet nd
S-wave phase shift~\cite{Bd98} and for higher partial-waves (in
both the quartet and doublet channels actually)~\cite{Ga00}.

\subsubsection{The tricky case: $\lambda > \lambda_{\rm c}$}

\begin{figure}[t]
\centerline{\psfig{file=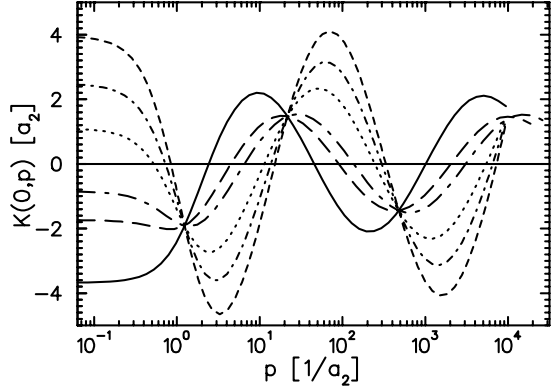,height=2.0in}}
\vskip -6 mm
 \caption{The amplitude $a(p)$ (denoted here $K(0,p)$), in the bosonic
case for different values of the momentum cutoff $\Lambda$ in units of
the two-body scattering length $a_2$. Figure from Ref.~\cite{Bd98},
courtesy U.~van Kolck.}
\label{fig-oscillations}
\end{figure}

The ``problem'' alluded to in the title of this subsection does
not arise until we consider couplings $\lambda$ sufficiently large
to generate imaginary solutions to Eq.~(\ref{eq:paulo}). In
particular, for the case originally of interest here, $\lambda=1$
the solution is:
\begin{equation}
s=\pm {\rm i} s_0 \, ,
\end{equation}
with $s_0=1.0064$. Then, in the intermediate domain between the
low-momentum scale $\gamma$ and the cutoff $\Lambda$ we anticipate
that the solution $a(p)$ will behave as:
\begin{equation}
a(p)=C\cos\left[s_0 \log\left({p \over \Lambda}\right) + \delta\right].
\end{equation}
Since there are now two admissible solutions to the integral
equation (\ref{eq:UV}) two parameters which are not determined by
the integral equation make an appearance in $a(p)$: $C$ and
$\delta$. The value of these parameters is ultimately fixed by
dynamics near the cutoff $\Lambda$. And this means that the phase
$\delta$ is fixed by ultraviolet, or short-distance, physics. The
result is that in this case the low-momentum behaviour of $a(p)$
varies wildly as the cutoff is changed.  (See
Fig.~\ref{fig-oscillations}.).  This is unacceptable from a
theoretical point of view: the cutoff is a theoretical artifice,
and dynamics near the cutoff should not play a key role in
determining the low-energy scattering in general and the
scattering length in particular.

One way to understand this difficulty is to realize that the
nucleon-exchange potential here can be thought of as a $1/r^2$
force law if we work in hyperspherical co-ordinates and consider
$1/\Lambda \ll r \ll 1/\gamma$.  This is a singular potential, and
the corresponding Hamiltonian is unbounded from below. The
difficulties in defining quantum-mechanical problems based on such
interactions are well-known~\cite{PP70,Be01}, since systems tend
to ``fall to the centre'' as they can always lower their energy by
cramming more of the wave function into the short-distance region.
Of course, this is not advantageous if we are dealing with a {\it
repulsive} $1/r^2$ potential, and so that problem is well-defined.
Now we gain some insight into why the nd quartet resulted in a
well-defined quantum mechanics problem. In that case the Pauli
principle precludes all three nucleons from getting close
together, and the result is a repulsive nucleon-exchange potential
between the deuteron and the third nucleon. In contrast, in the
triton, or in the three-boson system, the Pauli principle does not
forbid ``collapse to the centre'', and both the triton and the
trimer cannot be described by Eq.~(\ref{eq:aeq}), since that
equation fails to make a concrete prediction for $a_3$, or indeed,
for any low-energy scattering observable.

\subsection{The solution}

\begin{figure}[t]
\centerline{\psfig{file=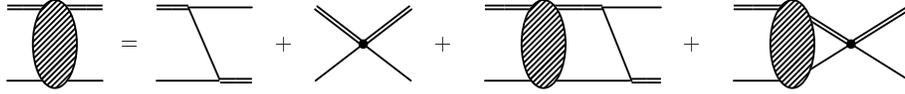,height=0.8in,
                 bbllx=60pt,bblly=690pt,bburx=560pt,bbury=770pt}}
\vskip -8 mm
\caption{Equation for the three-body amplitude including the three-body force.
Figure from Ref.~\cite{Be00}.}
\label{fig-3Bforcefadeev}
\end{figure}

The presence of such sensitivity to short-distance physics
suggests that the problem requires renormalization. For those used
to solving EFTs in perturbation theory this is surprising, since
the difficulty the Eq.~(\ref{eq:aeq}) does not arise at any finite
order in perturbation theory, it only occurs when we consider the
non-perturbative solution to the problem of neutron-deuteron
scattering~\cite{Bd99B,Ge00}.  However, in considering the
non-perturbative solution to Eq.~(\ref{eq:aeq}) it is perhaps not
surprising, given non-compact nature of the integral equation's
kernel in the limit $\Lambda \rightarrow \infty$, or,
equivalently, the singular behaviour of the one-nucleon-exchange
``potential''. Regardless of whether it is surprising or not, the
presence of the unphysical behaviour with variation in $\Lambda$
depicted in Fig.~\ref{fig-oscillations} suggests that we add to
the kernel a counterterm. A natural candidate is the three-body
interaction we already included in the Lagrangian
(\ref{eq:3Btransvestite}).  In retrospect, dropping the three-body
force term from the integral equation is not justified in the case
of bosons (although it is justified in S-wave quartet nd
scattering). Graphs containing the three body force are now
included as shown in Fig.~(\ref{fig-3Bforcefadeev}).

The new graphs depicted here modify the kernel of the integral
equation for threshold scattering to:
\begin{equation}\label{eq:Hkernel}
K(q,p)={1 \over 2qp}\,
 \log \left({p^2 + p q + q^2 + \gamma^2 \over p^2-p q + q^2 + \gamma^2} \right)
+ {H \over \Lambda^2},
\label{eq:K3B}
\end{equation}
where $H$ is related to the coefficient $h$ appearing in the
Lagrangian (\ref{eq:3Btransvestite}) by:
\begin{equation}
h={M H y^2 \over \Lambda^2},
\label{eq:notsonaive}
\end{equation}
so that the parameter $H$ is dimensionless.

For $H(\Lambda)\sim 1$ the three-body force has an effect in
Eq.~(\ref{eq:Hkernel}) only for $p$ in the vicinity of the cutoff.
Therefore the analysis of the behaviour of the amplitude given in
the previous subsection is still valid in the intermediate region
$\gamma \ll p \ll \Lambda$. It is now relatively straightforward
to choose $H=H(\Lambda)$ in such a way that the low-momentum part
of $a(p)$ is (essentially) independent of cutoff. In other words
we want to try and effect a choice of $H$ such that:
\begin{equation}
a(p)=C(\Lambda,H(\Lambda)) \cos\left[s_0 \log\left(\frac{p}{\Lambda}\right)
+ \delta(\Lambda,H(\Lambda))\right],
\end{equation}
contains a phase which is independent of the cutoff. To this end
we define an equation for the evolution of $H(\Lambda)$ with the
cutoff $\Lambda$ (a ``renormalization group'' equation):
\begin{equation}
-s_0 \log \Lambda + \delta(\Lambda,H(\Lambda))=-s_0 \log \Lambda_* \, ,
\end{equation}
where $\Lambda_*$ is some fixed physical scale.

\begin{figure}[t]
\centerline{\psfig{file=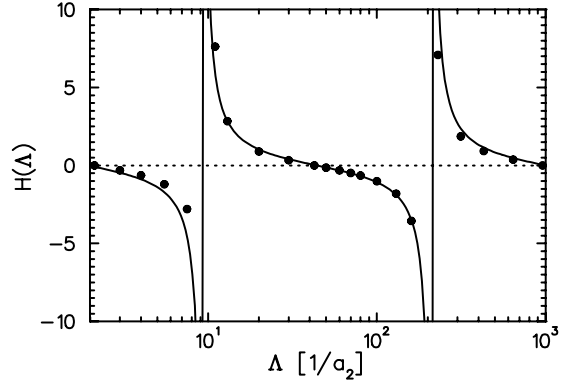,height=5 cm}}
\vskip -6 mm
\caption{The running of the ``three-body force'' $H(\Lambda)$ as a
function of $\Lambda$. The solid line is the prediction of Eq.~(\ref{eq:H})
and the points are values obtained for $H$ by adjusting it numerically
so as to reproduce a given three-body scattering length $a_3$. Figure from
Ref.~\cite{Bd99B} courtesy of U.~van Kolck.}
\label{fig-H}
\end{figure}

A fairly simple calculation (see Ref.~\cite{Bd99B}) then reveals
that an (approximate) solution to this is provided by:
\begin{equation}
\label{eq:H}
H(\Lambda)\ =\
- \ {\sin\left( s_0 \log\left[\Lambda/\Lambda_*\right]
                 - \arctan\left[1/s_0\right] \right)
\over
\sin\left( s_0 \log\left[\Lambda/\Lambda_* \right]
                 + \arctan\left[1/s_0\right] \right) }\, .
\end{equation}
Choosing $H$ as per this formula, then inserting it into the kernel
(\ref{eq:K3B}) and solving the integral equation for $a(p)$ should
ensure that the low-momentum behaviour of $a$ is relatively
insensitive to the cutoff $\Lambda$. This procedure can also be
implemented numerically, and the results agree startlingly well.
(See Figure~\ref{fig-H}.)

The key point here is the appearance of a new physical parameter
(e.~g. $\Lambda_*$ here) that is {\it not} determined by two-body
physics. This means that in systems such as the Helium trimer, or
the nd doublet channel, on-shell two-body scattering data alone is
not enough to fix the three-body observables. One additional
parameter is required, and this can be thought of as $\Lambda_*$
itself or the value of the three-body force $H(\Lambda_*)$ at that
scale.

\begin{figure}[t]
\centerline{\psfig{file=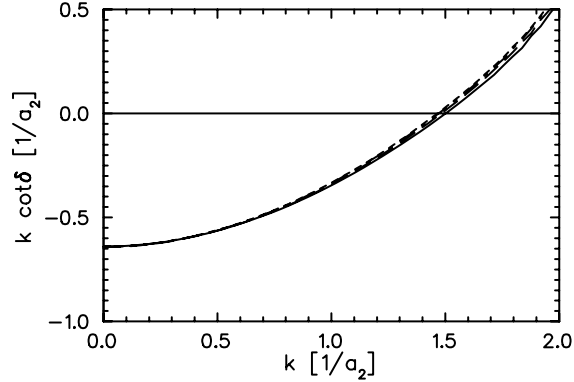,height=5 cm}}
\vskip -5 mm
\caption{Predictions for $k \cot \delta$ as a function of the relative
momentum of the dimer and helium atom $k$, in the case that $a_3=1.52
a_2$.  The different curves are for four different cutoffs ($\Lambda
a_2=42.6,100.0,230.0,959.0$), with $H$ adjusted in each case to ensure
that $a_3=1.52 a_2$. Figure from Ref.~\cite{Bd99B} courtesy of U.~van
Kolck.}
\label{fig-kcotd}
\end{figure}

Here I have not discussed the integral equation for the
half-off-shell scattering amplitude at finite energies. Sufficed
to say that this can be derived in exactly the same fashion as was
done for the zero-energy equation here. And the issue of the
low-momentum amplitude's sensitivity to the cutoff also appears at
any finite energy. However, in the region $k \ll p \ll \Lambda$
($k$ the on-shell momentum of the dimer and boson) the above
analysis applied to $a(p)$ goes through, since all reference to
the scale $k$ drops out. Consequently, {\it the renormalization of
the problem at $k=0$ through the addition of an energy-independent
$H(\Lambda)$ is sufficient to renormalize the problem at any
finite $k$ too}. (For a nice way of viewing this see
Ref.~\cite{HM01}.) Thus the energy-dependence of the phase shifts
is predictive. In boson-dimer scattering renormalizing in order to
reproduce the ratio $a_3/a_2$ given by Eqs.~(\ref{eq:a3}) and
(\ref{eq:a2}), yields the curve shown in Fig.~\ref{fig-kcotd} for
$k \cot \delta$ in S-waves. The curves shown are for different
choices of the cutoff, with $H(\Lambda)$ adjusted in each case to
give the same $\left.k \cot \delta\right|_{k=0}$.  Once this has
been done remaining effects of the regulator are suppressed by
$k^2/\Lambda^2$. These results are actually in accord with the
(scarce) experimental data on the trimer system, and also with
sophisticated model calculations of Helium atom-dimer
interactions.  The beauty of the EFT approach described here lies
in its simplicity, and its consequent ability to generate
universal predictions for energy-dependence of $k \cot \delta$
once the ratio $a_3/a_2$ is specified.

\begin{figure}[t]
\centerline{\psfig{file=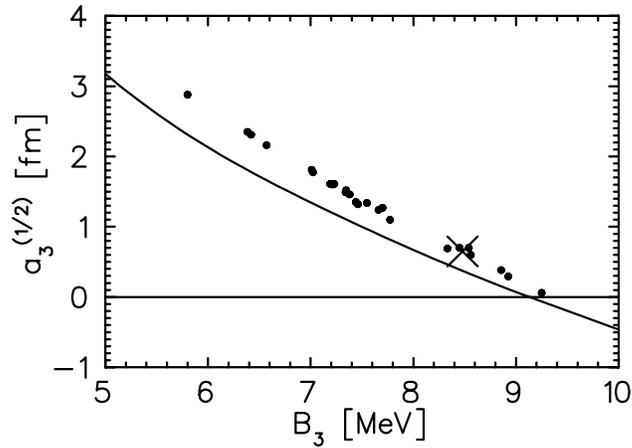,height=2.3in}}
\vskip -5 mm
\caption{Results for the doublet S-wave
neutron-deuteron scattering length for different models having very
similar two-body scattering lengths and effective ranges, denoted by
the dots.  The solid line is the EFT prediction. The cross indicates
the experimental value. Figure courtesy U. van Kolck, from
Ref.~\cite{Bd00B}.}
\label{fig-phillips}
\end{figure}

Such a universal prediction can be made for the relationship
between any two low-energy observables. For instance, once $a_3$
is fixed, the three-body bound-state energy $B_3$ is a prediction
of the theory. For a given choice $(\Lambda,H(\Lambda))$ we have a
definite prediction for both $a_3$ and $B_3$ and so we can
generate a parametric plot of the function $B_3(a_3)$. This is
what has been done in Fig.~\ref{fig-phillips}. In this case the
plot is for the case of the triton, and I will suppress the
additional spin-isospin complications that arise in this channel.
(For details see Ref.~\cite{Bd00B}.) I will just point out that
exactly the physics I have been discussing here results in the
one-parameter relationship between $B_3$ and $a_3$. The existence
of such a relationship is an old story in the nd
system~\cite{Ph68,AS73}~\footnote{The Phillips line has
traditionally been understood as arising from different off-shell
behaviours in the two-body NN amplitudes represented by the
different NN models~\cite{AS73}. In fact, this explanation is not
in conflict with the one given here, in which the Phillips line is
understood through the necessity for a three-body force, since
field redefinitions change both off-shell behaviour of amplitudes
and strengths of three-body forces~\cite{Fu00}.}. Plotting the
predictions of a number of different NN models which all give the
same low-energy NN data, we see that they give different
predictions for both $B_3$ and $a_3$. But these different
predictions are absolutely correlated: if we put all of them on a
graph of $B_3(a_3)$ then they fall on a straight line, known as
the Phillips line (not me, another Phillips).

The plot Fig.~\ref{fig-phillips} shows that the existence of this
Phillips line can be understood as being due to the dual facts:
\begin{itemize}
\item Two-body NN data is not sufficient to determine the low-energy
scattering in the doublet nd channel.

\item One, and only one, additional parameter, which can be
adjusted to one piece of NNN experimental data, then allows
concrete predictions to be made for low-energy NNN observables.
\end{itemize}

\section{Adding pions: a brief discussion of
using chiral perturbation theory for NN-system processes}
\label{sec-chipt}

We now wish to discuss energies larger than the very low energies
at which the effective field theory without explicit pion degrees
of freedom is valid. We will consider interactions of nucleons and
pions in the kinematic regime where
\begin{equation}
p \sim q \sim m_\pi \ll m_\rho, M
\label{eq:kindomain}
\end{equation}
with $p$ the (three)-momentum of the nucleons, $q$ the
(four)-momentum of any external probes, $m_\rho$ the rho-meson
mass and $M$ the nucleon mass.

In this kinematic domain an important constraint on pions'
interactions with each other and with nucleons is provided by the
spontaneously-broken approximate chiral symmetry of QCD. I will
not provide a review of the physics of chiral symmetry here.
Sufficed to say that extensions of the techniques described in
Section 1 can be used to derive an effective field theory in which
the low-energy degrees of freedom are nucleons and pions. The
symmetry which constrains their interactions is the approximate
$SU(2)_{\rm L} \times SU(2)_{\rm R}$ symmetry of QCD. But since
this symmetry is spontaneously broken and approximate the
development of this effective field theory requires that we also
understand how to introduce into the EFT operators which break
$SU(2)_{\rm L} \times SU(2)_{\rm R}$ in the same fashion in which
it is broken in QCD. The resultant Lagrangians, the development of
which is reviewed in Refs.~\cite{Br95,Le01,Me01}, and in the
lectures of Prof.~Leutwyler in this volume, represent an effective
quantum field theory valid in the kinematic domain
(\ref{eq:kindomain}). The resulting theory is known as chiral
perturbation theory ($\chi$PT). The small parameter in which we
expand is now
\begin{equation}
P \equiv {p \over \Lambda_{\chi \rm{SB}}}, {q \over \Lambda_{\chi\rm{SB}}},
{m_\pi \over \Lambda_{\chi \rm{SB}}},
\label{eq:Q}
\end{equation}
where $\Lcsb$ is the scale of chiral symmetry breaking:
\begin{equation}
\Lambda_{\chi \rm{SB}} \sim m_\rho, M, 4 \pi f_\pi.
\end{equation}

To summarize years of work on $\chi$PT in a (very) hand-waving
fashion, the two key ideas that make the $\chi$PT expansion
tenable are:
\begin{itemize}
\item That the short-distance effects of QCD which are not
strongly constrained by chiral symmetry can be expanded in powers
of the small momenta in the problem $p$ and $q$, with the
resulting operators having coefficients whose sizes are determined
by naive dimensional analysis, with the high-energy scale being
$\Lcsb$.

\item That the ``long-distance'' loop effects due to low-energy
pions and nucleons interacting involve only weak interactions,
since chiral symmetry implies that the chiral Lagrangian can be
organized in such a way that the pions only ever couple
derivatively to the nucleons and to each other. Thus every pion
interaction involves a factor of
\begin{equation}
{\partial_\mu \pi \over f_\pi},
\end{equation}
and so pionic loops are ultimately suppressed by factors of
\begin{equation}
{(k,m_\pi) \over 4 \pi f_\pi}
\end{equation}
where $k$ is the magnitude of the pion's three momentum.
\end{itemize}

We have already seen that the ``typical'' momentum scale relevant
for binding in deuterium is $\sqrt{MB}$ which is small compared to
$\Lcsb$.  Although the typical momentum in the trinucleons is
higher, it is still well below $\Lcsb$. Thus, we expect to be able
to calculate the response of such nuclei to low-energy probes
using this effective field theory. The result is a
systematically-improvable, model-independent description of
electroweak processes on these nuclei.  Furthermore, pionic
processes now fall within the purview of the EFT, and although I
will not discuss them here, calculations of reactions such as $\pi
{\rm d} \rightarrow \pi {\rm d}$ and $\gamma {\rm d} \rightarrow
\pi^0 {\rm d}$ in $\chi$PT have been quite
successful~\cite{Be97C,Be97,Be02}.

In this section I will discuss two recent calculations of
electromagnetic reactions on deuterium in chiral perturbation
theory that I have been involved in.  The reactions are elastic
electron-deuteron scattering and Compton scattering on deuterium.

\subsection{Power counting}

We begin by establishing the power counting for processes in this
approach.  While this power counting can be expressed
diagrammatically, \'a la that derived above for $\nopi$, here I
will lay out the power counting in the fashion of a simple quantum
mechanics problem. If you want to derive the diagrammatic rules
from the ones I give here I leave that to you as an exercise!

Consider an elastic scattering process on the deuteron whose
amplitude we wish to compute. If $\hat{O}$ is the transition
operator for this process then the amplitude in question is simply
$\langle \psi| \hat{O} |\psi \rangle$, with $|\psi \rangle$ the
deuteron wave function. Here I follow
Weinberg~\cite{We90,We91,We92}, and divide the formulation of a
systematic expansion for this amplitude into two parts: the
expansion for $\hat{O}$, and the construction of $|\psi \rangle$.

Chiral perturbation theory gives a systematic expansion for
$\hat{O}$ of the form:
\begin{equation}
\hat{O}=\sum_{n=0}^\infty \hat{O}^{(n)},
\label{eq:expansion}
\end{equation}
where we have labeled the contributions to $\hat{O}$ by their
order $n$ in the small parameter $P$ of Eq.~(\ref{eq:Q}).
Eq.~(\ref{eq:expansion}) is an operator statement, and the nucleon
momentum operator $\hat{p}$ appears on the right-hand side.
However, the only quantities which ultimately affect observables
are expectation values such as $\langle \psi| \hat{p} |\psi
\rangle$. For light nuclei this number is generically small
compared to $\Lambda_{\chi {\rm SB}}$.

To construct $\hat{O}^{(n)}$ one first writes down the vertices
appearing in the chiral Lagrangian up to order $n$. One then draws
all of the two-body, two-nucleon-irreducible, Feynman graphs for
the process of interest which are of chiral order $P^n$. The rules
for calculating the chiral order of a particular graph are:
\begin{itemize}
\item Each single-nucleon propagator scales like $1/P$, since in
(heavy-baryon) $\chi$PT the nucleon is a static object (at leading
order) and its propagator is just $1/p_0$, with $p_0$ the amount
by which the nucleon's four momentum is taken away from the
mass-shell by the process in question;

\item Each loop contributes $P^4$;

\item Graphs in which both particles participate in the reaction
acquire a factor of $P^3$;

\item Each pion propagator scales like $1/P^2$;

\item Each vertex from the $n$th-order piece of the chiral
Lagrangian contributes $P^n$.
\end{itemize}

These rules are consequences of naive dimensional analysis, pure
and simple. The argument is simply that the powers of $P$ derived
from considering the construction of $\hat{O}^{(n)}$ in this way
{\it must} be carried by the low-energy scales $p$, $q$, and
$m_\pi$. Meanwhile as the dimension of $\hat{O}^{(n)}$ increases
more and more powers of the high-energy scale $\Lcsb$ must appear
in the denominator in order for Eq.~(\ref{eq:expansion}) to be
dimensionally correct.

While this is a simple argument it has the significant consequence
that more complicated graphs contributing to $\hat{O}$, i.~e. ones
that involve two-body mechanisms, and/or higher-order vertices,
and/or more loops, are suppressed by powers of $P$.

\subsection{Deuteron wave functions}

There remains the problem of constructing a deuteron wave function
which is consistent with the operator $\hat{O}$. Weinberg's
proposal was to construct a $\chi$PT expansion in
Eq.~(\ref{eq:expansion}) for the NN potential $V$, and then solve
the Schr\"odinger equation to find the deuteron (or other nuclear)
wave function~\cite{We90,We91,We92}.  Recent calculations have
shown that nucleon-nucleon scattering data can be understood, and
deuteron bound-state static properties reliably computed, with
wave functions derived from $\chi$PT in this
way~\cite{Or96,KW97,Ep99,Re99,EM01}.

\begin{figure}[t]
\centerline{\epsfysize=3.5in \epsfbox{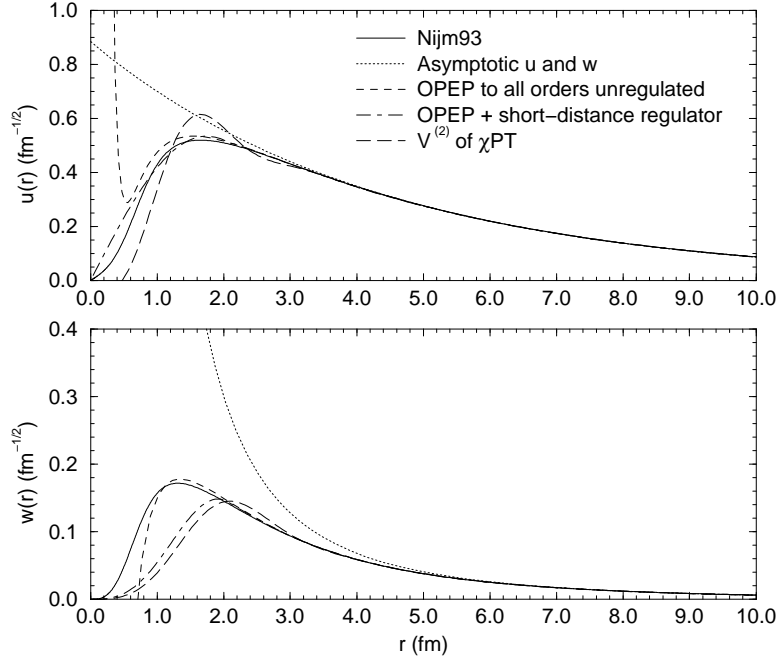}}
\vspace{-0.7cm}
\caption{Radial S-wave and D-wave wave deuteron
wave functions for several different NN potentials. The dotted
line is the ERT/$\nopi$ wave function, while the dot-dashed line
is that from a simple model with one-pion exchange at long
distances and a square well at short distances~\cite{PC99}. The
long-dashed line is the NLO deuteron wave function of
Ref.~\cite{Ep99} and the solid line is the Nijm93 wave
function~\cite{St94}.}
\label{fig-comparison}
\end{figure}

Now for $\chi$PT in the Goldstone-boson and single-nucleon sector
loop effects are generically suppressed by powers of the small
parameter $P$. This is just a consequence of the weakness of
pionic interactions in the chiral limit. Thus, in these reactions
the power counting in $P$ applies to the amplitude, and not to the
two-particle potential. However, the existence of the deuteron
tells us immediately that a power counting in which loop effects
are suppressed cannot be correct for the two-nucleon case, since
if it were we could use successive Born approximations to compute
the NN scattering amplitude, and there would be no NN bound state.
Weinberg's proposal to instead power-count the potential is one
response to this dilemma. However, its consistency has been
vigorously debated in the literature (see~\cite{vK99,Be00} for
reviews).  Recently Beane {\it et al.}~\cite{Be01} have brought
some clarity to this discussion, by showing that Weinberg's
proposal is consistent in the $\siii-\diii$ channel. In fact,
pionic interactions are not ``weak'' in this channel since even in
the chiral limit a (tensor) potential
\begin{equation}
V_{\rm T}(r) \sim {1 \over r^3},
\end{equation}
exists between the two nucleons. This is an (infinitely!) strong
interaction. Beane {\it et al.} propose that $\chi$PT be
reformulated in NN systems as an expansion {\it about the chiral
limit}. They have shown that this is a consistent proposal, and
that there are some signs that such an expansion will converge,
albeit not particularly rapidly.

Here I adopt a more ``lowbrow'' approach.  One way to understand
$\chi$PT power-counting for deuteron wave functions is to examine
the deuteron wave function in three different regions.
\begin{enumerate}
\item In the region $R \gg 1/m_\pi$ the deuteron wave function is
described solely by the asymptotic normalizations $A_{\rm S}$,
$A_D$, and the binding energy $B$. These quantities are
observables, in the sense that they can be extracted from phase
shifts by an analytic continuation to the deuteron pole. This
region is the one in which $\nopi$ gives the ``correct'' deuteron
wave function.

\item The second region corresponds to $R \sim 1/m_\pi$. Here pion
exchanges play a key role in determining the NN potential $V$,
and, associatedly, the deuteron wave functions $u$ and $w$. The
leading effect comes from iterated one-pion exchange---as has been
known for at least fifty years. Calculations with one-pion
exchange (OPE) defining the potential in this regime will be
referred to below as ``leading-order'' (LO) calculations for the
deuteron wave function. Corrections at these distances come from
two-pion exchange, and these corrections can be consistently
calculated in $\chi$PT. Two-pion exchange effects are suppressed
by powers of the small parameter $P$, with the ``leading''
two-pion exchange suppressed by $P^2$ relative to OPE. This
two-pion exchange can be calculated from vertices in ${\cal
L}_{\pi {\rm N}}^{(1)}$ and its inclusion in the NN potential
results in the so-called ``NLO'' calculation described in detail
in Ref.~\cite{Ep99}. Corrections to this two-pion-exchange result
from replacing one of the NLO two-pion-exchange vertices by a
vertex from ${\cal L}_{\pi {\rm N}}^{(2)}$. This results in an
additional suppression factor of $P$, or an overall suppression of
$P^3$ relative to OPE, and an ``NNLO'' chiral
potential~\cite{Or96,KW97,Ep99,Re99}.

\item Finally, at short distances, $R \ll 1/m_\pi$ we cutoff the
chiral one and two-pion-exchange potentials and put in some
short-distance potential whose parameters are arranged so as to
give the correct deuteron asymptotic properties. This cutoff will
generally be kept finite here. It remains an open question whether
it is useful to attempt to take $R \rightarrow 0$, in the manner
that we essentially did for $\nopi$.
\end{enumerate}

A plot of different deuteron wave functions is displayed in
Fig.~\ref{fig-comparison}.

\subsection{Elastic electron scattering on deuterium}

\begin{figure}[t]
\centerline{\epsfig{figure=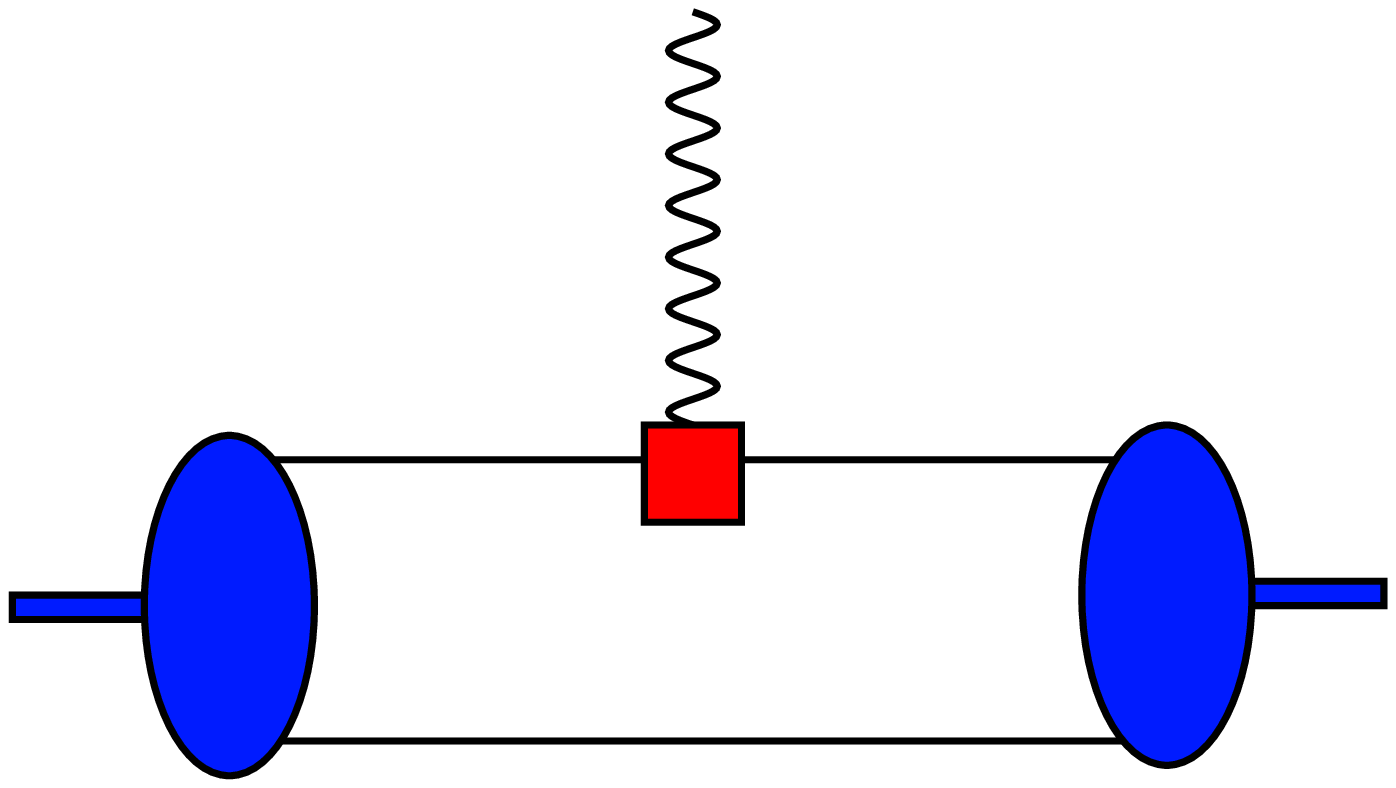,height=.15\textheight,width=.25\textwidth}
            \epsfig{figure=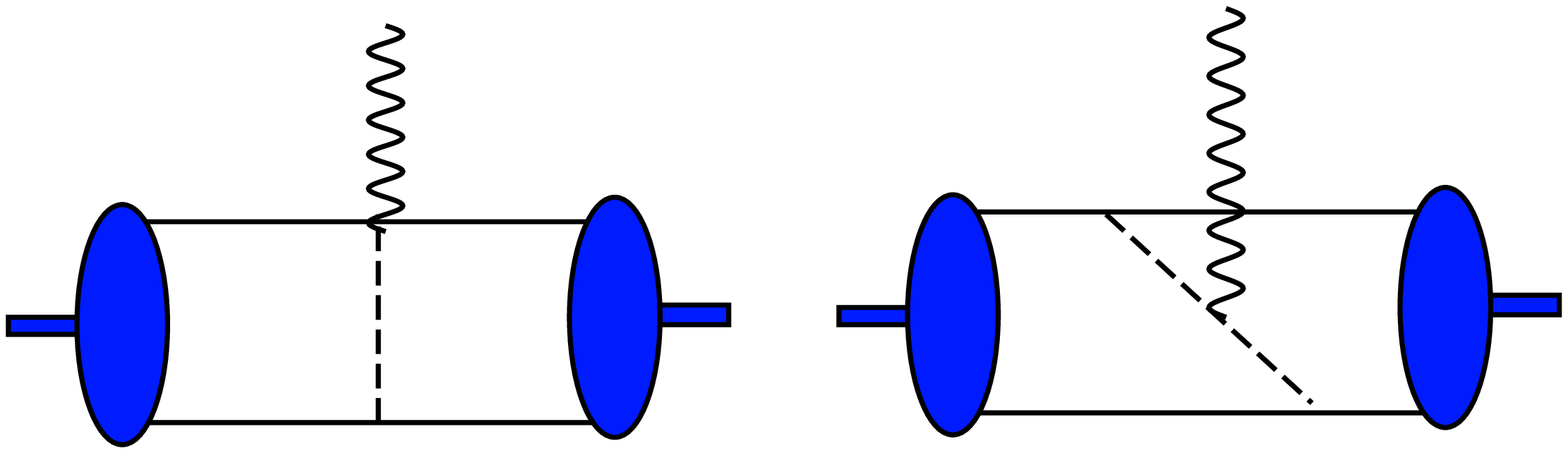,height=.15\textheight,width=.6\textwidth}}
\vspace{-0.6cm}
   \caption{The impulse-approximation contribution to $G_{\rm C}$ and $G_{\rm Q}$
    is shown on the left, while two meson-exchange current
    mechanisms which would contribute were the deuteron not an
    isoscalar target are depicted in the middle and on the right.}
\label{deuterongraphs}
\end{figure}

Both the deuteron's charge form factor, $G_{\rm C}$ (defined in
Eq.~(\ref{eq:GC})), and its quadrupole form factor, $G_{\rm Q}$, involve matrix
elements of the zeroth-component of the deuteron four-current,
$J^0$. Here we split $J^0$ into two pieces: a one-body part, and a
two-body part. The one-body part of $J^0$ begins at order $e$, with
the impulse approximation diagram calculated with the non-relativistic
single-nucleon charge operator for strutcutreless
nucleons. Corrections to the single-nucleon charge operator from
relativistic effects and nucleon structure are suppressed by two
powers of $P$, and thus arise at $O(eP^2)$, which is the
next-to-leading order (NLO) for $G_{\rm C}$ and $G_{\rm Q}$. At this
order one might also expect meson-exchange current (MEC)
contributions, such as those shown in
Fig.~\ref{deuterongraphs}. However, all MECs constructed with vertices
from ${\cal L}_{\pi N}^{(1)}$ are isovector, and so the first effect
does not occur until N$^2$LO, or $O(e P^3)$, where an $NN \pi \gamma$
vertex from ${\cal L}_{\pi N}^{(2)}$ replaces the upper vertex in the
middle graph of Fig.~\ref{deuterongraphs}, and produces an isoscalar
contribution to the deuteron charge operator. (This exchange-charge
contribution was first derived by Riska~\cite{Ri84}.)

\begin{figure}[t]
\centerline{\epsfysize=3.3in  \epsfbox{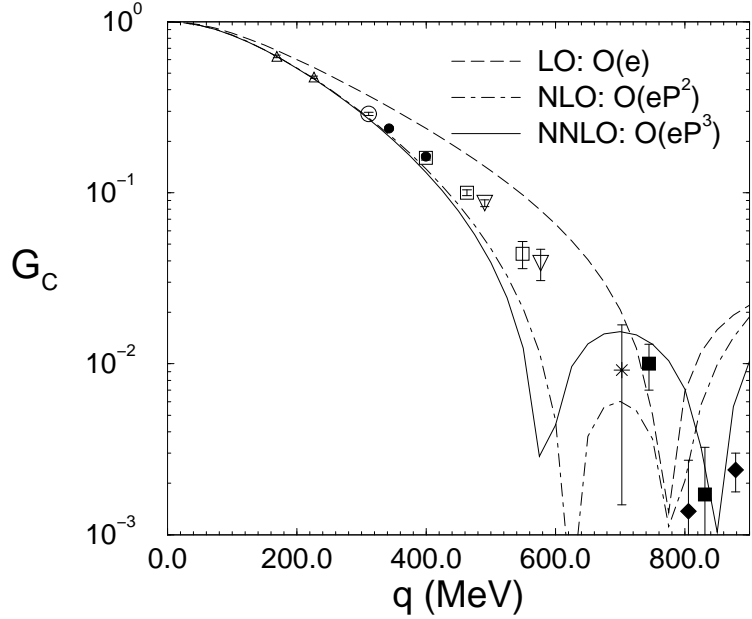} }
\vspace{-0.7cm}
\caption{The deuteron charge form factor at orders $e$, $eP^2$ and
$eP^3$ in a ``strict'' chiral perturbation theory calculation. The
experimental data is taken from the compilation of Ref.~\cite{Ab00B}.}
\label{fig-GCbad}
\end{figure}

The most important correction that arises at NLO [$O(eP^2)$] is the
inclusion of {\it nucleon} structure in $\chi$PT. At this order
isoscalar form factors are dominated by short-distance physics, and so
the only correction to the point-like leading-order result comes from
the inclusion of the nucleon's electric radius, i.e.
\begin{equation}
{G_{\rm E}^{\rm (s)}}_{\mbox{$\chi$PT NLO}}=1 - \frac{1}{6}
 \langle r_{\rm E}^{{\rm (s)} \, 2}
\rangle q^2.
\end{equation}
This description of nucleon structure breaks down at momentum
transfers $q$ of order 300 MeV. There is a concomitant failure in the
description of ed scattering data (see Figure~\ref{fig-GCbad}).  (This
failure was already seen in the work of Mei\ss ner and
Walzl~\cite{MW01}.). In order to focus on {\it deuteron} structure, in
the results presented below I have chosen to circumvent this issue by
using a ``factorized'' inclusion of nucleon
structure~\cite{Ph01}. This facilitates the inclusion of
experimentally-measured single-nucleon form factors in the
calculation, thereby allowing us to test how far the theory is able to
describe the {\it two-body} dynamics that takes place in ed
scattering.

\begin{figure}[t]
\centerline{\epsfysize=3.8in  \epsfbox{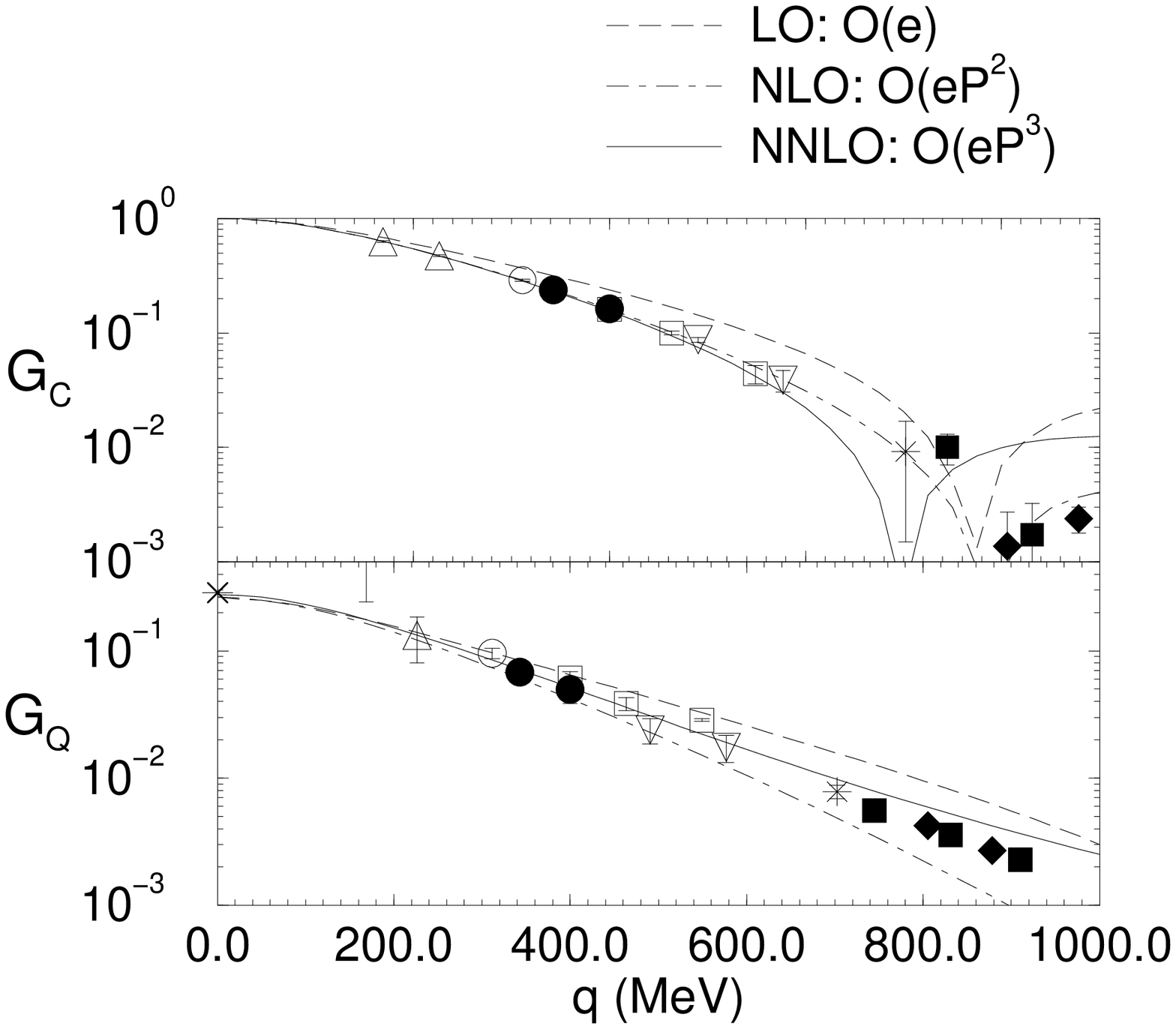} }
\vskip -5 mm
\caption{The deuteron charge and quadrupole form factors to order
$eP^3$ in chiral perturbation theory, with the effects of nucleon
structure included in the calculation. The experimental data is
taken from the compilation of Ref.~\cite{Ab00B}. $G_{\rm Q}$ is in
units of fm$^2$. }
\label{fig-FCFQ}
\end{figure}

We now compute results for $G_{\rm C}$ and $G_{\rm Q}$ at LO
(structureless nucleons), NLO (adding nucleon structure and the first
relativistic corrections), and N$^2$LO (including the meson-exchange
contribution discussed above in the charge operator $J_0$).  This
yields the parameter-free predictions shown in
Fig.~\ref{fig-FCFQ}. The figure demonstrates that convergence is quite
good below $q \sim 600$ MeV---especially for $G_{\rm C}$. The results
shown are for the NLO chiral wave function, but the use of the NNLO
chiral wave function, or indeed of simple wave functions which include
only one-pion exchange, do not modify the picture greatly below
$q=700$ MeV~\cite{Ph01}. It is also clear that---provided the
experimental data from electron-nucleon scattering is taken into
account---$\chi$PT is capable of describing the charge and quadrupole
form factors of deuterium at least as far as the minimum in $G_{\rm
C}$. This result is encouraging for the application of EFT to light
nuclei.

\subsection{Compton scattering on deuterium}
\label{sec-gammagammad}

\noindent The scattering of low-energy photons from the nucleon at
low energies probes the long-distance structure of the neutron and
the proton.  In the case of the neutron the low-energy Hamiltonian
for the neutron's interactions with photons is exactly
(\ref{eq:Hatomnat}). The coefficients of the ${\bf E}^2$ and ${\bf
B}^2$ terms are related to the electric ($\alpha_{\rm n}$) and magnetic
($\beta_{\rm n}$) polarizability of the neutron via~\cite{Hl00}:
\begin{equation}
a_1=-2 \pi \alpha_{\rm n}; \quad a_2=-2 \pi \beta_{\rm n}.
\end{equation}

Photon-nucleon scattering has been studied in $\chi$PT in
Ref.~\cite{Br92,Br92B}, where the following results for these
neutron polarizabilities were obtained at leading non-vanishing
order ($O(e^2 P)$):
\begin{eqnarray}
\alpha_{\rm n}={{5 e^2 g_{\rm A}^2}\over{384 \pi^2 f_\pi^2 m_\pi}}
 =12.2 \times 10^{-4} \, {\rm fm}^3;
 ~~\beta_{\rm n}={{e^2 g_{\rm A}^2}\over{768 \pi^2 f_\pi^2 m_\pi}}=
 1.2 \times 10^{-4} \, {\rm fm}^3 \, .
\end{eqnarray}
Chiral perturbation theory also predicts $\alpha_{\rm n}=\alpha_{\rm p}$,
$\beta_{\rm n}=\beta_{\rm p}$ at this order.
The processes which contribute at $O(e^2P)$ are those in which the
photon interacts with the nucleon's pion cloud. Recent
experimental values for the proton polarizabilities
are~\cite{To98}:
\begin{eqnarray}
\alpha_{\rm p} + \beta_{\rm p}=13.23 \pm 0.86^{+0.20}_{-0.49} \times 10^{-4} \,
{\rm fm}^3, \nonumber\\
\alpha_{\rm p} - \beta_{\rm p}=10.11 \pm 1.74^{+1.22}_{-0.86} \times 10^{-4} \,
 {\rm fm}^3 \, ,
\label{polexpt}
\end{eqnarray}
where the first error is a combined statistical and systematic
error, and the second set of errors comes from the dispersion
relation analysis used to extract $\alpha_{\rm p}$ and $\beta_{\rm
p}$ from the experimental data.  These values are in good
agreement with the $\chi$PT predictions.

Meanwhile, the neutron polarizabilities $\alpha_{\rm n}$ and
$\beta_{\rm n}$ are difficult to measure, due to the absence of
free neutron targets, thus this prediction is not well tested.
Hence there has been a considerable effort to extract $\alpha_{\rm
n}$ and $\beta_{\rm n}$ by performing Compton scattering on
nuclear targets. Coherent Compton scattering on a deuteron target
has been measured at $E_\gamma=$ 49 and 69 MeV by the Illinois
group \cite{Lu94} and $E_\gamma= 84.2-104.5$ MeV at
Saskatoon~\cite{Ho00}.

However, the amplitude for Compton scattering on the deuteron
involves mechanisms other than Compton scattering on the
individual nucleons inside the deuterium nucleus. Hence, if we
want to extract neutron polarizabilities we will need a
theoretical calculation of Compton scattering on the deuteron in
which we can make sure that the error in our extraction of
$\alpha_{\rm n}$ and $\beta_{\rm n}$ from subtracting out these
``two-body'' mechanisms is under control. $\chi$PT is an ideal
framework for this, since the two-body mechanisms that contribute
to $\gamma {\rm d} \rightarrow \gamma {\rm d}$ are calculable
order-by-order in the chiral expansion, and so we can perform a
consistent extraction of $\alpha_{\rm n}$ and $\beta_{\rm n}$. The
theoretical error is defined as the size of the mistake made by
neglecting terms of the next order in the expansion.

\begin{figure}[t]
\centerline{\epsfysize=2.8in  \epsfbox{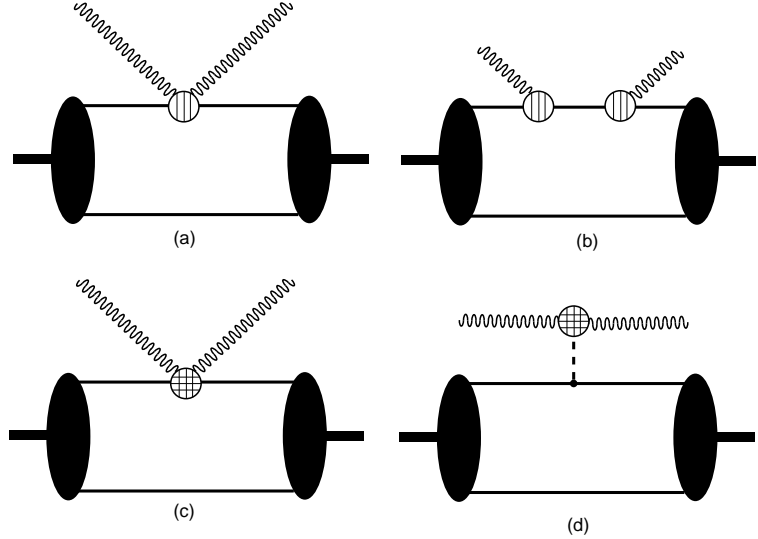} }
\vskip -5 mm
   \caption{Graphs
   which contribute to Compton scattering on the deuteron at
   ${\cal O}(e^2)$ (a) and ${\cal O}(e^2P)$ (b-d).  The
   sliced and diced blobs are from ${\cal L}_{\pi {\rm N}}^{(3)}$ (c) and
   ${\cal L}_{\pi \gamma}^{(4)}$ (d).  Crossed graphs are not
   shown.}
\label{procomptonebodmod}
\end{figure}

\begin{figure}[b]
\centerline{\epsfysize=1.6in  \epsfbox{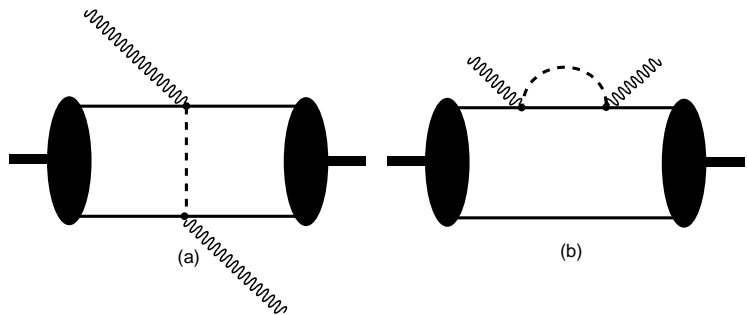} }
\vskip -5 mm
\caption{Graphs which contribute to Compton scattering
on the deuteron at ${\cal O}(e^2P)$.
Crossed graphs are not shown.}
\label{combined}
\end{figure}

The Compton amplitude we wish to evaluate is (in the $\gamma d$
center-of-mass frame):
\begin{eqnarray}
T^{\gamma {\rm d}}_{M' \lambda' M \lambda}({\bf k}',{\bf k})&=&
\int {{\rm d}^3p \over (2 \pi)^3} \, \,
\psi_{M'}\left( {\bf p} + {{\bf k} - {\bf k}' \over 2}\right)
 \, \, T_{\gamma {\rm N}_{\lambda' \lambda}}({\bf k}',{\bf k})
 \, \, \psi_M({\bf p})\nonumber\\
&+& \int {{{\rm d}^3p \, \, {\rm d}^3p'} \over (2 \pi)^6}
\, \, \psi_{M'}({\bf p}') \, \,
T^{\rm 2N}_{\gamma {\rm NN}_{\lambda' \lambda}}({\bf k}',{\bf k})
\, \, \psi_M({\bf p})
\label{eq:gammad}
\end{eqnarray}
where $M$ ($M'$) is the initial (final) deuteron spin state, and
$\lambda$ ($\lambda'$) is the initial (final) photon polarization
state, and ${\bf k}$ (${\bf k}'$) the initial (final) photon
three-momentum, which are constrained to $|{\bf k}|=|{\bf
k}'|=\omega$.  The amplitude $T_{\gamma {\rm N}}$ represents the
graphs of Fig.~\ref{procomptonebodmod} and Fig.~\ref{combined}b
where the photon interacts with only one nucleon. These are the
``one-body'' mechanisms we actually wish to investigate.
Meanwhile, the amplitude $T^{\rm 2N}_{\gamma {\rm NN}}$ represents
the graphs of Fig.~\ref{combined}a where there is an exchanged
pion between the two nucleons. These are the ``two-body''
mechanisms we wish to calculate in a controlled way and then
subtract them off.

The LO contribution to Compton scattering on the deuteron is shown
in Fig.~\ref{procomptonebodmod}(a).  This graph involves a vertex
from ${\cal L}_{\pi {\rm N}}^{(2)}$ and so is $O(e^2)$.  This
contribution is simply the Thomson term for scattering on the
proton. There is thus no sensitivity to either two-body
contributions {\it or} nucleon polarizabilities at this order.  At
$O(e^2 P)$ there are several more graphs with a spectator nucleon
(Figs.~\ref{procomptonebodmod}(b),(c),(d)), as well as graphs
involving an exchanged pion with leading order vertices
(Fig.~\ref{combined}(a)) and one-loop graphs with a spectator
nucleon (Fig.~\ref{combined}(b))~\cite{Be99}. Graphs such as
Fig.~\ref{combined}(b) contain the physics of the proton and
neutron polarizabilities at $O(e^2P)$ in $\chi$PT.

\begin{figure}[t]
\centerline{\epsfig{figure=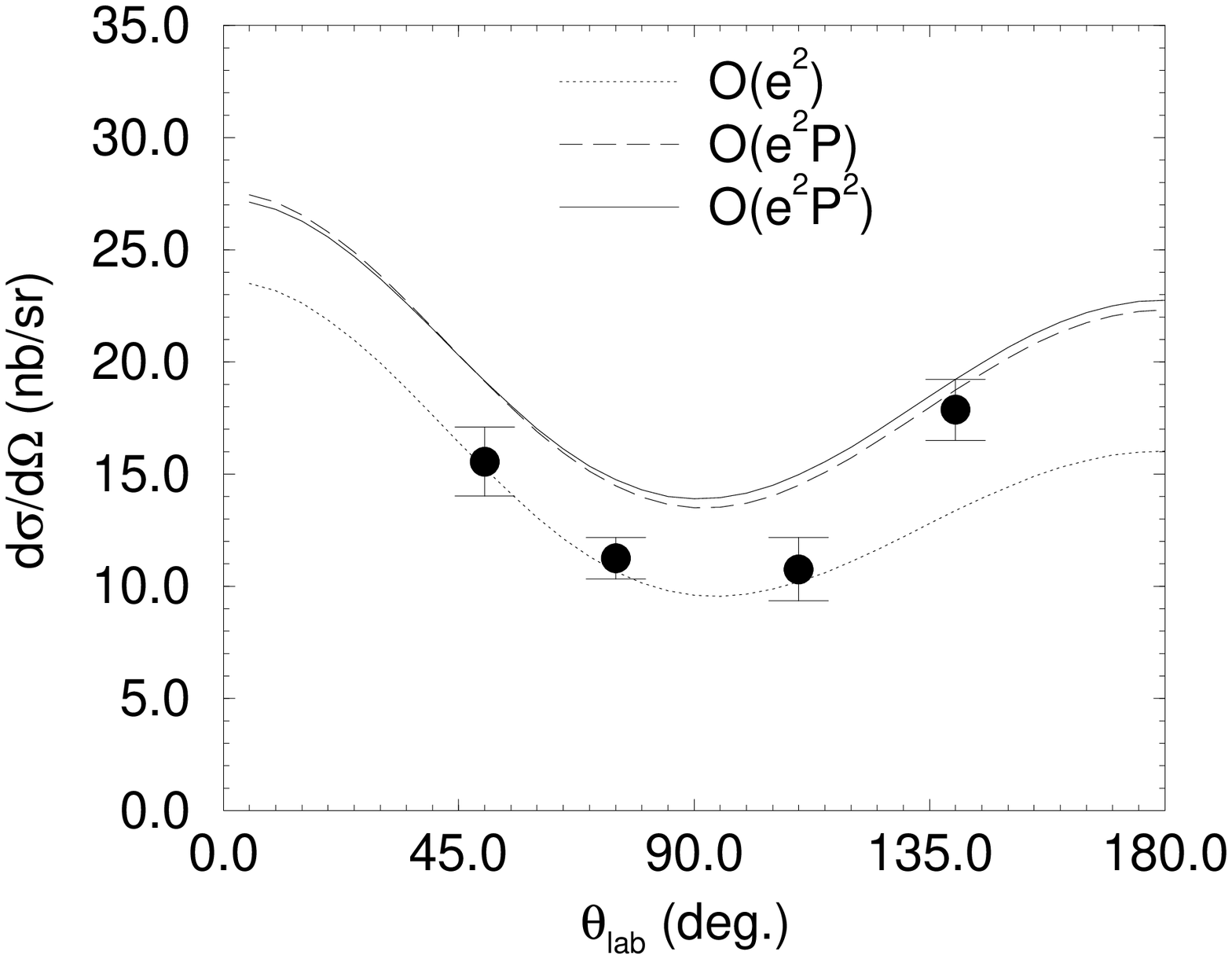,height=.2\textheight,width=.3\textwidth}
            \epsfig{figure=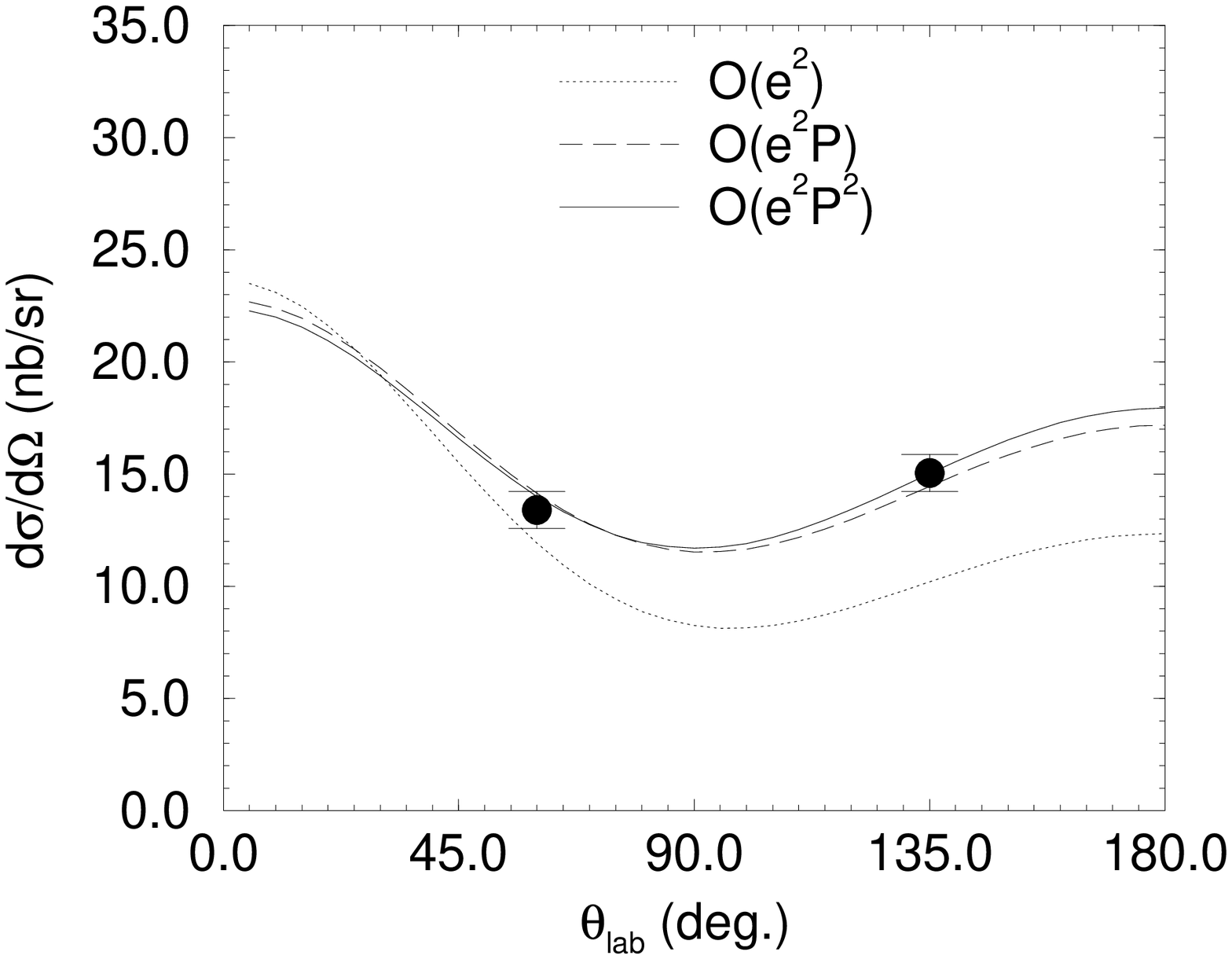,height=.2\textheight,width=.3\textwidth}
            \epsfig{figure=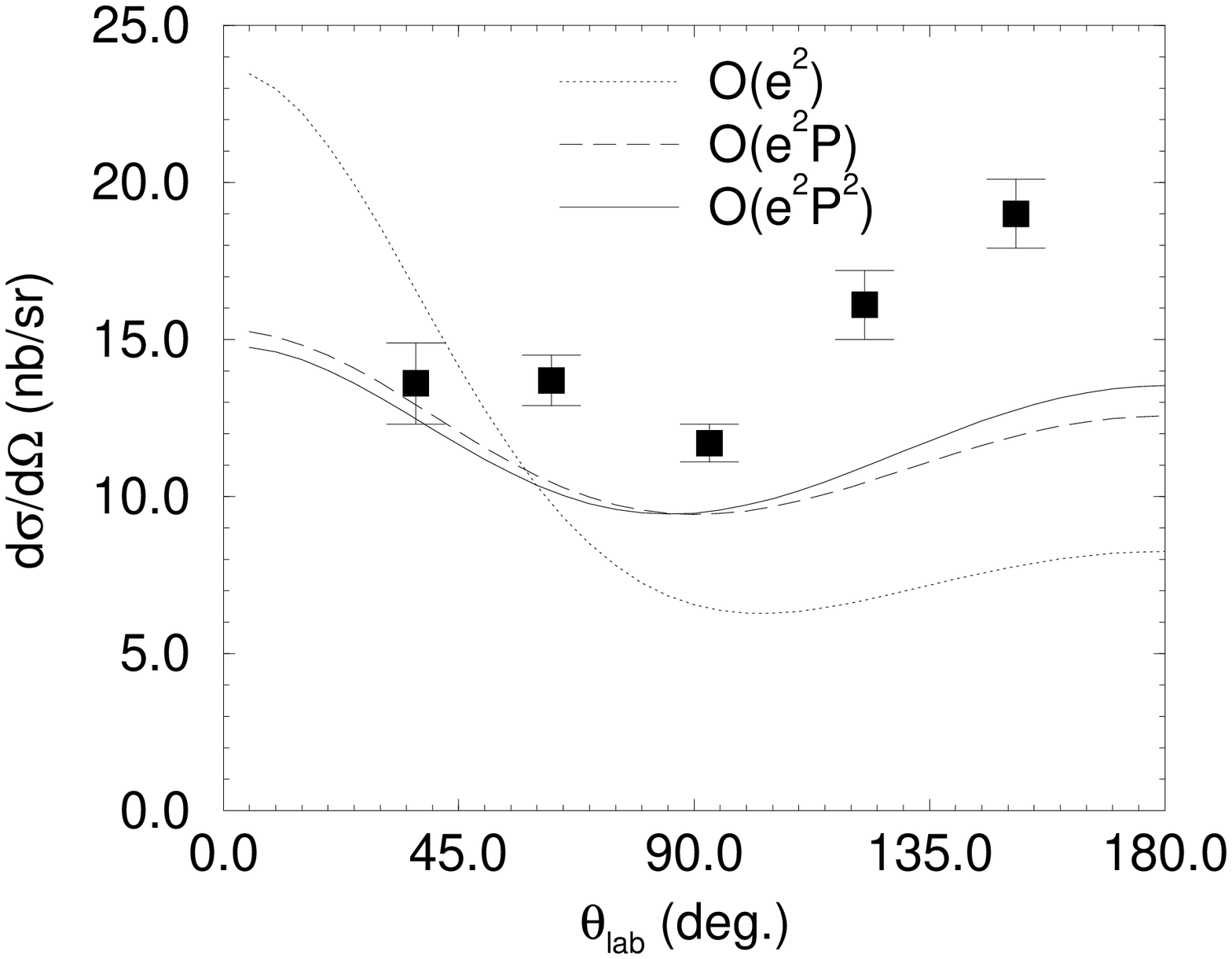,height=.2\textheight,width=.3\textwidth}}
\vskip -5 mm
\caption{Results of the $O(e^2)$ (dotted line), $O(e^2P)$ (dashed
line), and $O(e^2 P^2)$ (solid line) calculations for $E_\gamma=49~{\rm
MeV}$, $69~{\rm MeV}$ and $95~{\rm MeV}$ respectively from left to
right.}
\label{cvgplot}
\end{figure}

The results shown below were generated with the NLO chiral wave
function of Ref.~\cite{Ep99}.  Fig.~\ref{cvgplot} shows the
results at $E_\gamma = $ 49, 69, and 95 MeV.  For comparison we
have included the calculation at $O(e^2)$ in the kernel, where the
second contribution in Eq.~(\ref{eq:gammad}) is zero, and the
single-scattering contribution is given solely by
Fig.~\ref{procomptonebodmod}(a).  At $O(e^2P)$ all contributions
to the kernel are fixed in terms of known pion and nucleon
parameters, so to this order $\chi$PT makes {\it predictions} for
deuteron Compton scattering.

We also show the $O(e^2 P^2)$ result from work just being
completed~\cite{Be02}. This calculation uses the NLO wave function
of Ref.~\cite{Ep99} and the $O(e^2 P^2)$ single-nucleon Compton
amplitudes derived in Ref.~\cite{McG01}. Also included are a
number of $1/M$ corrections to the two-body piece of the $\gamma
{\rm NN} \rightarrow \gamma {\rm NN}$ amplitude. It turns out that
such mechanisms are the only corrections to the two-body piece of
this amplitude at this amplitude. In other words, no two-body
counterterms contribute to $\gamma {\rm d} \rightarrow \gamma {\rm
d}$ at $O(e^2 P^2)$.  On the other hand, at $O(e^2 P^2)$
single-nucleon counterterms which shift the polarizabilities enter
the calculation. But that is an advantage!  It is exactly these
counterterms---or more specifically the neutron
counterterms---that we wish to extract from the $\gamma {\rm d}$
data, since these counterterms serve to fix the value of
$\alpha_{\rm n}$, $\beta_{\rm n}$, $\alpha_{\rm p}$ and
$\beta_{\rm p}$.  In this sense Compton scattering on deuterium at
$O(e^2 P^2)$ is analogous to the reaction $\gamma {\rm d}
\rightarrow \pi^0 {\rm d}$~\cite{Be97}: an $O(e^2 P^2)$
calculation allows us to test the single-nucleon physics which is
used to predict the results of coherent scattering on deuterium,
since there are no undetermined parameters in the two-body
mechanisms that enter to this order in the chiral expansion. For
the calculations displayed here we have {\it set} the nucleon
polarizabilities to the values:
\begin{eqnarray}
\alpha_{\rm p}=\alpha_{\rm n}=12.0 \times 10^{-4}~{\rm fm}^3,\\
\beta_{\rm p}=\beta_{\rm n}=2.0\times 10^{-4}~{\rm fm}^3.
\end{eqnarray}
In fact the results for this observable are quite sensitive to
$\alpha_{\rm n}$ and $\beta_{\rm n}$, so ultimately one would like to
use experimental values for the proton polarizabilities, such as those
quoted in Eq.~(\ref{polexpt}), as well as performing a $\chi^2$ fit to
the photon-{\it deuteron} experimental data in order to constrain the
neutron $\alpha$ and $\beta$. This should become a more realistic
option as data on $\gamma {\rm d} \rightarrow \gamma {\rm d}$ from Lund becomes
available~\cite{Lu02}.

The data shown in Fig.~\ref{cvgplot} includes six points from the
Illinois experiment at 49 and 69 MeV~\cite{Lu94} and the Saskatoon
data at 95 MeV~\cite{Ho00}. Statistical and systematic errors have
been added in quadrature. It is quite remarkable how well the
$O(e^2)$ calculation reproduces the 49 MeV data. However, since
there are sizeable higher-order corrections this agreement might
be regarded as somewhat fortuitous~\footnote{In fact, the
``Weinberg'' power counting used here breaks down at these lower
photon energies, since it is designed for $\omega \sim m_\pi$, and
does {\it not} recover the deuteron Thomson amplitude as $\omega
\rightarrow 0$.  Correcting the power counting to remedy this
difficulty appears to improve the description of the 49 MeV data,
without significantly modifying the higher-energy
results~\cite{Be99}.}. Meanwhile, the agreement of the $O(e^2P)$
calculation with the 69 MeV data is very good, although only
limited conclusions can be drawn. Finally, the results at 95 MeV
disagree markedly with the Saskatoon data, especially at backward
angles. While this could be indicative that the neutron
polarizabilities are starkly incorrect tests suggest that
modifying $\alpha_{\rm n}$ and $\beta_{\rm n}$ in order to
reproduce the backward-angle $\gamma {\rm d}$ data from Saskatoon
would lead to disagreement between our calculations and the 69 MeV
data.

In fact, similar conclusions have been reached in potential-model
calculations~\cite{Wl95,LL98,KM99} where the results for the
differential cross sections are comparable to those obtained here
using $\chi$PT. They are also not dissimilar to those obtained in
NN EFTs without explicit pions (see Ref.~\cite{GR01}).

Many other processes have been computed using Weinberg
power-counting. The application to $\pi {\rm d}$ scattering is
intriguing, and holds promise of allowing for an extraction of the
isoscalar $\pi {\rm N}$ scattering length~\cite{Be97C,We92}. The
prediction of the threshold amplitude for $\gamma {\rm d} \rightarrow
\pi^0 {\rm d}$ was a significant triumph for $\chi$PT applied to
deuterium~\cite{Be97}. Meanwhile, Park, Min, Kubodera, Rho, and
collaborators have calculated many processes, paying particular
attention to astrophysical reactions. See Ref.~\cite{Be00,Pa00B}
for summaries of this work, which represents an important
application of $\chi$PT in particular, and effective theories of
nuclear physics in general.

\section*{Acknowledgments}

I thank the organizers of the International Summer School for inviting
me to present these lectures there, and for hosting a stimulating
summer school in the beautiful city of Prague. I also want to thank my
EFT collaborators and interlocutors: Iraj Afnan, Silas Beane, Paulo
Bedaque, Michael Birse, Tom Cohen, Harald Grie\ss hammer, Hans Hammer,
Christoph Hanhart, Judith McGovern, Manuel Malheiro, Ulf Mei\ss ner,
Gautam Rupak, Martin Savage, Roxanne Springer, and Bira van Kolck, for
many enjoyable and educative discussions about effective field
theories applied to nuclear physics. Thanks also to Vladimir
Pascalutsa and Bira van Kolck for comments on the manuscript, to
Vincent Stoks and Evgeni Epelbaum for data on their wave functions,
and to Jacques Ball for information on the parameterization of
experimental data on $G_{\rm C}$ and $G_{\rm Q}$ for deuterium.  This
work was supported by the U.~S. Department of Energy under grant
number DE-FG02-93ER40756.

\end{document}